\documentclass[graybox,oribibl]{svmult}

\usepackage{type1cm}        % activate if the above 3 fonts are                 
\usepackage{makeidx}         % allows index generation       
\usepackage{graphicx}
\usepackage{multicol}        % used for the two-column index                    
\usepackage[bottom]{footmisc}% places footnotes at page bottom                  
\usepackage{newtxtext}       %       
\usepackage[varvw]{newtxmath}       % selects Times Roman as basic font 

\usepackage{amsmath,dsfont}
\usepackage{braket}
\usepackage{color}

\makeindex             % used for the subject index                             
\newcommand{\Fkt}[1]{\,\mathrm {#1}}

\ifx\Tr\renewcommand{\Tr}{\Fkt{Tr}}
\else\newcommand{\Tr}{\Fkt{Tr}}
\fi
\ifx\tr\renewcommand{\tr}{\Fkt{tr}}
\else\newcommand{\tr}{\Fkt{tr}}
\fi

\newcommand{\openone}{\mathds{1}}

\usepackage[style=phys,biblabel=brackets]{biblatex}
\usepackage[colorlinks]{hyperref} % no boxes around links
\usepackage{hypernat} %enable collapsing of citations even if
\hypersetup{%
   pdftitle={Introduction to quantum control \today},
   pdfsubject={Varenna school lecture notes},
   pdfauthor={Christiane Koch},
   pdfstartpage=1,
   pdfstartview=FitH,
   linkcolor=blue,
   filecolor=blue,      
   urlcolor=cyan,
   citecolor=blue,
 }
\usepackage{orcidlink}
 
\begin{document}

\title*{Introduction to quantum control: From basic concepts to applications in  quantum technologies}
\titlerunning{Quantum control: From basic concepts to applications}
\author{Christiane~P.~Koch~\orcidlink{0000-0001-6285-5766}}

\institute{Christiane P. Koch \at 
Freie Universität Berlin,
Dahlem Center for Complex Quantum Systems \& Fachbereich Physik, Arnimallee 14, 14195 Berlin, Germany, \email{christiane.koch@fu-berlin.de}}
\maketitle

\abstract{% 164 words
  Quantum control refers to our ability to manipulate quantum systems. This tutorial-style chapter focuses on the use of classical electromagnetic fields to steer the system dynamics. In this approach, the quantum nature of the control stems solely from the underlying dynamics, through the exploitation of destructive and constructive interference to reach the control target. We first discuss two basic control principles --- coherent control which uses manipulation in frequency or time to design these interferences, and  adiabatic following where access to the control target is enabled by tracking the time-dependent ground state. For complex control targets and system dynamics that exceed the scope of these basic principles, optimal control theory provides a powerful suite of tools to design the necessary protocols. A key consideration for the successful application of optimal control theory is a proper choice of the optimization functional. All concepts are illustrated using recent work from my research group, with a focus on controlling atoms and superconducting qubits. The chapter concludes with an outlook on integrating coherent control with engineered dissipation and a discussion of open questions in the field.
}

\section{Introduction}
\label{sec:intro}

The superposition principle captures the essence of quantum mechanics and manifests itself in quantum interference. In textbook examples, these interferences are typically taken as given, arising from static interactions. Remember the dihydrogen molecule where the difference between the 
bonding and antibonding orbitals is just the sign, i.e., a relative phase of $\pi$, in the respective superpositions of atomic orbitals. 
Coherent quantum control involves a change in perspective: Interferences can be created dynamically. The interaction of a quantum system with an external field allows for creating superpositions and manipulating the relative phases in these superpositions. Under this angle, quantum control appears as a way to see the superposition principle at work.

Quantum control also refers to our ability to manipulate quantum systems in the desired way. It is a field at the intersection of physics, mathematics, chemistry, and engineering. From a practical viewpoint, it is not necessary to think about superpositions. Rather, quantum control can be seen as an engineering discipline, where the equations of motion are given by physics (or chemistry). Using the language of mathematics, it  provides a set of tools to implement tasks that arise in the operation of quantum computers or quantum sensors. In this mindset, quantum control is at its best when it can be used in a black box way. While fully justified when the goal is to design and build a quantum device, the engineering perspective falls short of harnessing the exploratory power of quantum control.

The present lecture attempts a more holistic view, uniting perspectives from physics, chemistry, mathematics, and engineering. Such an interdisciplinary approach establishes a constructive cycle for quantum control: Analyzing the requirements for operating a quantum device leads to the design of new algorithms; implementing these algorithms yields practical protocols; and analyzing the underlying control mechanism improves our microscopic understanding. The lecture is structured as follows: 
Section~\ref{sec:basic} will introduce the basic concepts, starting from the statement of the control problem in Sec.~\ref{subsec:problem}. It provides an overview over the theoretical tools that are nowadays available to solve quantum control problems -- from the basic principles of coherent control in frequency and time in Sec.~\ref{subsec:coherent} and control based on adiatic evolution in Sec.~\ref{subsec:adiabatic} to optimal control theory in Secs.~\ref{subsec:oct} and \ref{subsec:functionals} and controllability analysis in Sec.~\ref{subsec:controllability:basics}. The application of these tools will be illustrated in Sec.~\ref{sec:applications} with selected examples, with a focus on the control of open quantum systems and a strong bias towards work from my group (a reader interested in a more comprehensive summary of the state of the art is referred to Refs.~\cite{GlaserEPJD15,KochEPJQT22}).
After discussing the design of suitable figures of merit for open quantum systems in Sec.~\ref{subsec:open:figure-of-merit}, the discussion is structured according to the control strategies for open quantum systems. Control avoiding  decoherence is covered in Sec.~\ref{subsec:open:fight},  followed by a discussion of quantum reservoir engineering as a strategy that exploits  the environment in Sec.~\ref{subsec:open:exploit}. As an intermezzo, Sec.~\ref{subsec:controllability} presents advances in controllability analysis, as a tool to select suitable interactions for the control of quantum reservoir engineering. A recent addition to the set of control strategies, explained in Sec.~\ref{subsec:open:create}, is the use of auxiliary quantum degrees of freedom in order to engineer desired dissipation. 
The lecture concludes  in Sec.~\ref{sec:open} with a discussion of open questions and possible avenues for future research, the selection of which again reflects a  personal preference.

\section{Basic concepts and methods}
\label{sec:basic}

\subsection{Statement of the control problem}
\label{subsec:problem}

The term ``control'' is widely used in quantum physics and often with rather diverse meanings. In this lecture, we take the perspective of \textsl{control theory} in the mathematical sense; that is, we ask how a dynamical system can be steered from a given initial state to a desired final state using one or several external control knobs. Here, the notion of dynamical system implies that we can write down (and also solve, either analytically or numerically) the system's equation of motion.

This formalization of the control problem is of course not limited to quantum physics. In fact, optimal control theory (where the control problem is solved using variational calculus to find a minimum or maximum~\cite{PontryaginBook}) has been developed to steer the motion of objects that follow the classical laws of motion. A famous example is the landing of a space craft on the moon. For simplicity, we assume that we can describe the space craft as a point particle. In this case, the equation of motion is given by Newton's law,
\[
m\,\ddot{\vec{r}}=\vec F\,,
\]
and the initial and final states are specified in terms of the point particle's initial, respectively desired, final position and velocity,
\[
  \vec r(t=0)=\vec r_0,\; \vec v(t=0)=v_0\,\quad\quad\quad
  \vec r(t=t_f)=\vec r_f,\; \vec v(t=t_f)=0\,,
\]
where the latter ensures a soft landing of the space craft. Control on the space craft is exerted by burning fuel which can be described by the corresponding force, $\vec F_{burn}$ and which also leads to a time-dependent mass $m(t)$. In this very simplified picture of a space craft's motion, the optimal control problem can be solved in closed form~\cite{OCTbook}.

%\cmt{show here cartoon for the three examples (spacecraft, molecules, qubit)?}

One of the areas of quantum physics, where control theory has been employed early on, is the control of photochemical reactions, %% add citation
i.e., the breaking (and much later also the making~\cite{LevinPRL15}) of chemical bonds in molecules using laser light. For molecules in the gas phase, the interaction with the laser typically occurs on such short timescales that the dynamics is well described by Schr\"odinger's equation,
\[
  \frac{\partial\ket{\psi}}{\partial t}=-i\hat H(t) \ket{\psi}\quad\quad\quad
  \ket{\psi(t=0)}=\ket{\psi_0},\, \ket{\psi(t=t_f)}=\ket{\psi_{target}}\,.
\]
Here and throughout we use units such that $\hbar =1$. $\ket{\psi_0}$ and $\ket{\psi_{target}}$ describe the molecule's initial and desired final state. The laser being the external control knob, the Hamiltonian comprises both the model for the molecule, $\hat H_0$ (including electronic and possibly also vibrational and rotational degrees of freedom), and the molecule's interaction with the laser. The laser-molecule interaction is usually well described in the electric dipole approximation such that the total Hamiltonian is given by
\begin{equation}\label{eq:H}
\hat H(t) = \hat H_0 + \hat H_I(t)=\hat H_0 + \hat{\vec d}\cdot\vec E(t)\,,    
\end{equation}
where $\hat{\vec d}$ is the molecule's dipole moment and $\vec E(t)$ the electric field. In many experiments, more than a single initial state is populated; in which case Schr\"odinger's equation needs to be solved for each of these states. Expectation values are then obtained as an incoherent average over all the states with their respective weights~\cite{HegerPRE19}, for example the Boltzmann weights for a thermal initial state. This is not just a technicality but important for the observability of quantum control: Typically, control leverages the superposition principle of quantum mechanics to realize the desired dynamics and thus gets reduced the less pure the initial state is~\cite{LevinPRL15}.

The purity %$Tr\left(\hat\rho^2_0\right)$
with which initial states can be prepared, together with the accuracy with which we know the Hamiltonian, is what makes the field of quantum information science an ideal playground for quantum control~\cite{KochEPJQT22}. At the same time, the requirements on the fidelity, i.e., the accuracy with which the desired dynamics is realized, are very stringent. Moreover, there is often no clear separation between the timescales of decoherence and the desired dynamics.
In order to find control schemes that realize the best possible fidelity, it is then necessary to include decoherence and describe the system state by $\hat\rho$. A common description for such dynamical systems uses the GKLS master equation~\cite{BreuerBook,LidarLectureNotes}, named after Gorini, Kossakowski, Lindblad and Sudarshan~\cite{Chruscinski2017}, 
\begin{equation}\label{eq:GKLS}
  \frac{\partial\hat\rho}{\partial t}= -i[\hat H,\hat\rho]_-+\mathcal L_D(\hat\rho)
  = -i[\hat H,\hat\rho]_-+\sum_j \left(\hat L_a\hat\rho\hat L_a^+
    -\frac{1}{2}\hat L_a^+\hat L_a\hat\rho-\frac{1}{2}\hat\rho\hat L_a^+\hat L_a
    \right)
  \,,
\end{equation}
where the dissipative part of the evolution is generated by $\mathcal L_D$. The jump operators $\hat L_a$ describe processes leading to loss of energy and phase, either phenomenologically or derived from a microscopic picture of the system's interaction with its environment~\cite{BreuerBook,LidarLectureNotes}.
Depending on the control target (as will be discussed in more detail below), we have to consider a single initial state, $\hat\rho(t=0)=\hat\rho_0$, or a set thereof, $\hat\rho_{0,j}$, $j=1,\ldots,M$, and similarly for the final state(s), $\hat\rho_{target,j}$, $j=1,\ldots,M$. Unless the external drives are very strong~\cite{AlbashNJP2012,DannPRA2018}, they enter only the Hamiltonian and not $\mathcal L_D$.

For the simple example of a qubit, in the most general case 
\begin{equation}
  \label{eq:Hqubit}
\hat H(t) = \frac{\omega_0}{2}\hat\sigma_z + u_1(t)\hat\sigma_z+u_2(t)\hat\sigma_x\,,  
\end{equation}
and the jump operators are $\hat L_1=\sigma^-$, $\hat L_2=\sigma^+$, describing decay, and $\hat L_3=\sigma_z$ for pure dephasing.
Recall that the coherent dynamics of a driven two-level system in the rotating-wave approximation (RWA, see Sec.~\ref{subsec:frameTLS}) can be solved analytically --- these are the famous Rabi oscillations~\cite{Cohen-TannoudjiQM2}: 
%\cmt{show picture of Rabi oscillations?}
The frequency of the oscillations is determined by the strength of the external control and its detuning from resonant driving. For zero detuning, the populations of the two levels oscillate between zero and one, as long as the external control is applied. This gives rise to control via so-called $\pi$- and $\pi/2$-pulses, the workhorses of many coherent spectroscopies,
where the duration of the external control is chosen such that the populations are inverted, respectively brought from a single level into an equal superposition of both levels. The control problem for a two-level system can  in many cases be solved analytically even with explicit account of decoherence~\cite{BoscainPRXQ2021}, using the techniques of optimal control theory~\cite{PontryaginBook}.

Given the formalization of the control problem in terms of control target and equations of motion, we can now look at ways to solve it. We will start with two intuitive approaches --- coherent control which leverages interference to implement the desired dynamics, and adiabatic following where the system state slowly transforms into the desired one, before presenting formal optimal control theory. The latter is generally applicable (provided we can solve the dynamics, analytically or numerically) but may yield solutions which are not always easy to interpret. In contrast, the first two are somewhat limited in their applicability, but provide clear insight into the control mechanism. They can serve as a starting point for further studies but also help interpret numerical optimal control results.

\subsection{Coherent control in frequency and time}
\label{subsec:coherent}

In Bohr's famous gedankenexperiment, an electron, when sent through a double-slit, interferes with itself. If a detector were to record the electron, it would yield an interference pattern where constructive and destructive interference take turns, depending on the distance between the slits. Nowadays such experiments are routinely carried out in the lab, with a grating instead of a double-slit and molecules rather than electrons~\cite{HornbergerRMP2012}. 
Even though these experiments are carried out with many molecules, it is always a single molecule that interferes with itself. The idea of a quantum object interfering with itself due to wave-particle duality can be made to work also in a more abstract sense, and coherent control takes it from the real space surrounding us to mathematical Hilbert space. Remarkably, the consequences are measurable, just as the interference pattern behind a double slit (or grating).

In coherent control, interferences are created between different "quantum pathways" connecting initial and final states~\cite{ShapiroBook,TannorBook}. These quantum pathways correspond to time evolutions due to the action of an external field. For example, in an atom, electric dipole transitions excited resonantly by the absorption of one photon or non-resonantly by the absorption of three photons, can connect the same initial and final states, and varying the relative phase between these two pathways modifies the population in the final state~\cite{ShapiroBook}. Because of the two laser colors involved, this is sometimes referred to as ``bichromatic control''. In a molecule without inversion symmetry, i.e., a chiral molecule, electric dipole selection rules also allow for the interference between one-photon and two-photon pathways~\cite{GoetzPRL19}. 

\begin{figure}[tbp]
    \centering
    \includegraphics[width=0.75\linewidth]{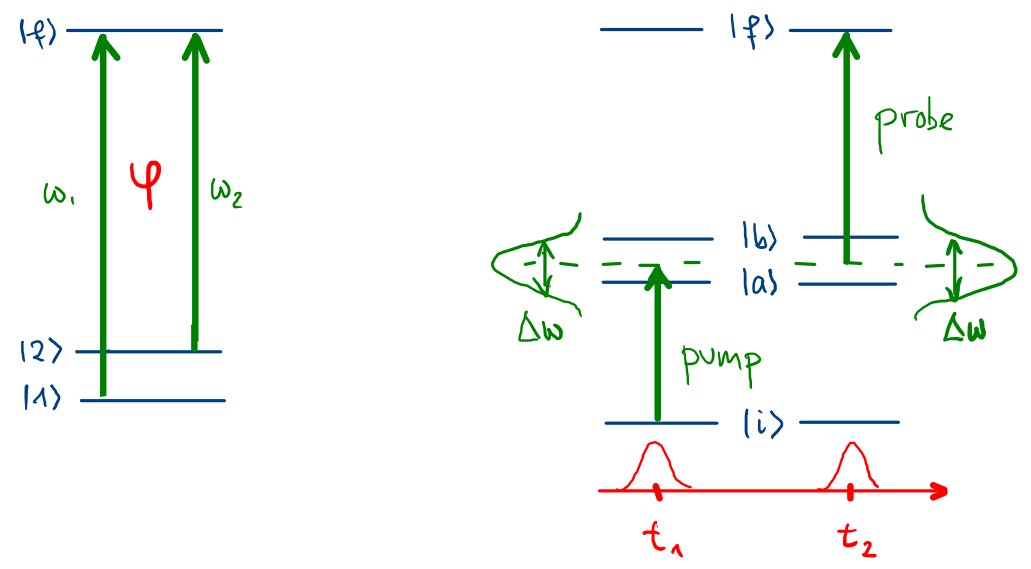}
    \caption{\textbf{Coherent control via quantum pathway interference.} Left: Spectral interference, starting from an initial superposition state (left). It is the laser phase $\varphi$ that controls the interference and hence the final population in $\ket{f}$. Right: Temporal interference where a pump pulse creates a superposition state which is subsequently further excited to a common final state by a probe pulse. It is the time delay between pump and probe pulse which controls the interference and hence the final population in $ket{f}$.}
    \label{fig:CC}
\end{figure}

The simplest example to see the interference~\cite{ShapiroBook}, depicted in Fig.~\ref{fig:CC} (left), assumes the initial state to be given by a coherent superposition of two levels, $\ket{\psi(t)}=c_1^0\ket{1}+c_2^0\ket{2}$, $c_{1,2}^0\in\mathbb{C}$, that are excited to the same final state, $\ket{f}$. A two-color laser field resonantly excites these two transitions, $\vec E(t)=\vec E_1\cos(\omega_1t)+\vec E_2\cos(\omega_2t+\varphi)$ where $\varphi$ denotes the relative laser phase.
We will describe the process within first order time-dependent perturbation theory~\cite{ShapiroBook}, 
\begin{equation}\label{eq:PT1}
    \ket{\psi^{(1)}(t)}=\frac{1}{i} \int_{t_0}^t dt' e^{-i H_0(t-t')} H_I(t') e^{-i H_0(t'-t_0)} \ket{\psi(t_0)}\,.
\end{equation}
Since the initial state is a superposition of $\ket{1}$ and $\ket{2}$, the only non-zero component of the final state in first order perturbation theory is $\langle f\ket{\psi^{(1)}(t_f)}=c_f^{(1)}(t_f)$.
Given the Hamiltonian, Eq.~\eqref{eq:H}, with $H_0=\mathsf{diag}(E_1,E_2,E_f)$ and expressing the interaction with the two-color laser field in terms of Rabi frequencies $\Omega_{ij}=\vec d_{fi}\cdot\vec E_{j}$ where $\vec d_{fi}$ denotes the electric dipole transition matrix element from level $i$ to $f$, Eq.~\eqref{eq:PT1} becomes
\begin{eqnarray*}
    c_f^{(1)}(t) &=& 
\frac{\Omega_{11}}{i}\int_{t_0}^t dt' e^{-i E_f(t-t')} \cos(\omega_1 t') e^{-i E_1(t'-t_0)} c_1(t_0)
\\&&+
\frac{\Omega_{12}}{i}\int_{t_0}^t dt' e^{-i E_f(t-t')} \cos(\omega_2 t'+\varphi) e^{-i E_1(t'-t_0)} c_1(t_0)
\\&&+
\frac{\Omega_{21}}{i}\int_{t_0}^t dt' e^{-i E_f(t-t')} \cos(\omega_1 t') e^{-i E_2(t'-t_0)} c_2(t_0)\\&&+
\frac{\Omega_{22}}{i}\int_{t_0}^t dt' e^{-i E_f(t-t')} \cos(\omega_2 t'+\varphi) e^{-i E_2(t'-t_0)} c_2(t_0)\,.
\end{eqnarray*}
Introducing $\tilde c^0_i=e^{-i E_f t}e^{i E_i t_0} c_i(t_0)$ and $\omega_{fi}=E_f-E_i$, we obtain
\begin{eqnarray*}
    c_f^{(1)}(t) &=& 
\frac{\Omega_{11}}{2i} \tilde c_1^0 \int_{t_0}^t dt' \left(e^{i (\omega_{f1}+\omega_1)t')} +e^{i (\omega_{f1}-\omega_1)t')} \right) + \ldots
\end{eqnarray*}
The terms that oscillate with the sums $\omega_{fi}+\omega_j$ of the transition and laser frequencies describe highly off-resonant transitions and are neglected within the rotating wave approximation, 
\begin{eqnarray*}
     c_f^{(1)}(t) &=& 
\frac{\Omega_{11}}{2i} \tilde c_1^0 \int_{t_0}^t dt' e^{i (\omega_{f1}-\omega_1)t')} +
\frac{\Omega_{12}}{2i} \tilde c_1^0 e^{-i\varphi}\int_{t_0}^t dt' e^{i (\omega_{f1}-\omega_2)t')}\\&& +
\frac{\Omega_{21}}{2i} \tilde c_2^0 \int_{t_0}^t dt' e^{i (\omega_{f2}-\omega_1)t')} +
\frac{\Omega_{22}}{2i} \tilde c_2^0 e^{-i\varphi}\int_{t_0}^t dt' e^{i (\omega_{f2}-\omega_2)t')}\,.
\end{eqnarray*}
Finally, we assume that the interaction with the laser field is much longer than any other timescale. This allows us to extend the integral boundaries to $\pm\infty$ such that we can replace the integrals by $2\pi\delta(\omega_{fi}-\omega_j)$. We then obtain for the final state amplitude
\[
c_f^{(1)} = -i\pi\left(\Omega_{11}\tilde c_1^0 +\Omega_{22}\tilde c_2^0 e^{i\varphi}\right)\,,
\]
since we had assumed the laser frequencies $\omega_{1/2}$ to match the two transition frequencies. Note that the result of the transition $\ket{1}\to\ket{f}$ being excited only by the $\omega_1$-component and not the $\omega_2$-component of the field (and conversely for $\ket{2}\to\ket{f}$) can also be obtained by first going into the interaction picture or a rotating frame and then invoking the (two-photon) rotating wave approximation~\cite{ShoreBook2011}, see App.\ref{subsec:frameNLS}). The interference shows up in the population, i.e., in the quantity that is measured:
\begin{equation}
    |c_f^{(1)}(t\to\infty)|^2=\pi^2\bigg(\Omega_{11}^2|\tilde c_1^0 |^2 +\Omega_{22}^2|\tilde c_2^0 |^2 + 2\mathfrak{Re}\left(\Omega_{11}\Omega_{22}(\tilde c_1^0)^*\tilde c_2^0 e^{-i\varphi}\right)
    \bigg)\,.
\end{equation}
More precisely, it is the last term that can take positive or negative values, corresponding to constructive or destructive interference. The interference depends on the values $\tilde c_1^0$, $\tilde c_2^0$, i.e., the initial populations and eigenenergies, as well as the relative phase of the laser field $\varphi$. It is the latter that can easily be tuned in an experiment and is thus 
the main control knob.

Bichromatic control, and its generalizations to more than two quantum pathways, see e.g. ref.~\cite{GoetzPRL19}, create the desired interference spectrally. A perfectly equivalent way is a temporal perspective~\cite{TannorBook}. Let us again assume that we start with a coherent superposition of two levels. In its most general form, the state is written 
\[
\ket{\psi(t)}= c_a(t)e^{-iE_at}\ket{a} + c_e(t)e^{-iE_bt}\ket{b}\,.
\]
The natural time evolution $e^{-iE_it}$ of the two levels is irrelevant, when populations are measured. If one can, however, directly probe the superposition, it gives rise to a time-dependent relative phase which results in the so-called "quantum beats". We can again use first order perturbation theory to see it. Let's assume we apply a weak probe field with a finite spectral width, and thus also finite temporal width i.e., a pulse, that excites both $\ket a$ and $\ket b$ to a common final state $\ket f$. According to perturbation theory, the population in $\ket f$ is proportional to $|\braket{f|\hat d|\psi(t)}|^2$. Assuming for simplicity an equal superposition, this results in 
\[
P_f(t) \sim |\braket{f|\hat d|\psi(t)}|^2 = \frac{1}{2}|d_{fa}|^2+\frac{1}{2}|d_{fb}|^2+
|d_{fa}|\cdot|d_{fb}|\cos((E_b-E_a)t)\,.
\]
Again, the interference shows up in the last term --- due to the sign change in the cosine, the last term can cancel or augment the other two terms, resulting in oscillations between zero and maximal population $P_f$. 

The simplest control scheme realizing coherent control in time is pump-probe spectroscopy, illustrated in Fig.~\ref{fig:CC} (right): A first pulse, or "pump", creates the superposition state $\ket{\psi(t)}$, and a second pulse probes the time evolution of the superposition by excitation to the common final state. Note that this control protocol is necessarily time-dependent: A single pulse (with small to moderate pulse intensity) can create a superposition state only when it has a spectral bandwidth $\Delta\omega>0$ ensuring  that all states of the superposition are resonantly excited. The spectral bandwidth translates into a finite pulse duration $\tau$, for example, $\tau=1/(\pi\Delta\omega)$ for transform-limited (unshaped) Gaussian pulses. Pump-probe control is widely used across different timescales and various physical systems, from attosecond pulses for electronic motion in molecules~\cite{KrauszRMP2024} to Ramsey interferometry in quantum sensing~\cite{DegenRMP2017}, with the common point being that the pulse duration must be much shorter than the dynamical timescale to be probed. 

It is important to note that while we have used perturbation theory to elucidate the control mechanism, coherent control based on spectral or temporal interference is not limited to weak fields. In fact, the stronger the fields, the more pathways will be excited that can interfere with each other~\cite{AmitayPRL2008}. Even outside of its validity regime, perturbation theory can be a very useful tool to rationalize control mechanisms since the book-keeping of the different interactions with the external field allow for easy identification of the interfering quantum pathways. A complementary perspective, relevant for strong external fields, is provided by analyzing the dynamics in terms of field-dressed states. These are at the core of control based on adiabatic following. 

\subsection{Adiabatic following}
\label{subsec:adiabatic}

The starting point for control via adiabatic following is the observation that the Hamiltonian $\hat H(t)$ can be diagonalized at any given instant of time $t$. The instantaneous eigenstates are also called ``dressed states'', as opposed to the bare states, i.e., the eigenstates of $\hat H_0$. 
While both eigenvalues and eigenvectors of $\hat H(t)$ depend on time, one can seek to stay in the same eigenstate, let's say the ground state. This is possible if the time-dependence of $\hat H(t)$ is sufficiently slow, since then the eigenstates for all $t$ are adiabatically connected. The condition for adiabaticity can be derived from the unitary transformation $\hat U(t)$ that diagonalizes $\hat H(t)$. Moreover, identifying which terms violate adiabaticity allows for designing control protocols that mitigate this violation. They are known as counterdiabatic driving~\cite{DemirplakJPCA2003} or shortcuts to adiabaticity~\cite{GueryOdelinRMP2019}. We will discuss this using again the example of a (strongly driven) two-level system.

Assume we have found the unitary $\hat U(t)$ that diagonalizes the Hamiltonian $\hat H(t)$ at time $t$,
\[
\hat U(t) \hat H(t) \hat U^+(t) = \hat D(t)\,,
\]
where $\hat D(t)$ denotes the diagonal matrix of the time-dependent eigenvalues. We can then use $\hat U(t)$ to transform the time-dependent Schr\"odinger equation into the instantaneous eigenbasis of $\hat H(t)$ (see App.~\ref{app:prerequ} for a reminder on frame transformations),
\begin{equation}\label{eq:adia}
i\frac{\partial}{\partial t}\ket{\tilde\psi(t)} = 
\left(\hat D(t)-i\hat U(t)\frac{\partial \hat U^+(t)}{\partial t}\right)\ket{\tilde\psi(t)}\,.    
\end{equation}
If the system starts in an eigenstate $\ket{\phi_n(t=0)}$ of $\hat H(t=0)$ and the dynamics is perfectly adiabatic, the state will evolve according to  $\exp(-i\lambda_n(t)t)$, where $\lambda_n(t)$ is one of the entries of $\hat D(t)$. In other words, the second term on the right-hand side of Eq.~\eqref{eq:adia} is only relevant beyond adiabatic evolution, it 
captures the non-adiabatic (or "diabatic") transitions! Adiabaticity can be ensured if this term is negligible compared to the first one. We will see below for the example of the two-level system that this requires large Rabi frequencies, i.e., large field amplitudes, and long and smooth pulses. An alternative is provided by counterdiabatic driving which adds further external controls to cancel the second term and thus effectively  suppress non-adiabatic transitions~\cite{DemirplakJPCA2003}.  

Moving to the two-level system with general Hamiltonian 
\begin{equation}\label{eq:HTLS}
\hat H(t)=\begin{pmatrix}
    E_a & W_{ab}(t)\\
    W^*_{ab}(t) & E_b
\end{pmatrix}    \,,
\end{equation}
the unitary that diagonalizes $\hat H(t)$ is given by
\begin{equation}
  \label{eq:UTLS}
\hat U(t) = \begin{pmatrix}
    \cos\frac{1}{2}\vartheta e^{i\frac{1}{2}\varphi} & \sin\frac{1}{2}\vartheta e^{-i\frac{1}{2}\varphi}\\ \sin\frac{1}{2}\vartheta e^{i\frac{1}{2}\varphi}
    & \cos\frac{1}{2}\vartheta e^{-i\frac{1}{2}\varphi}
\end{pmatrix}\,,  
\end{equation}
where the two angles are defined via
\[
\tan\vartheta=\frac{2|W_{ab}|}{E_a-E_b}\,,\quad W_{ab}=|W_{ab}|e^{i\varphi}\,.
\]
This yields for the diabatic transition matrix in Eq.~\eqref{eq:adia}
\begin{equation}\label{eq:dia}
i \hat U(t)\frac{\partial \hat U^+(t)}{\partial t} = \frac{1}{2}
\begin{pmatrix}
    -\frac{\partial\varphi}{\partial t} & i\frac{\partial \vartheta}{\partial t}\\
    i\frac{\partial \vartheta}{\partial t} & \frac{\partial\varphi}{\partial t}
\end{pmatrix}    \,.
\end{equation}
The diagonal terms in Eq.~\eqref{eq:dia} add a non-adiabatic phase to the time evolution but do not induce transitions between eigenstates $\ket{\phi_n(t)}$, $\ket{\phi_m(t)}$. It is the off-diagonal term, i.e., the time derivative of $\vartheta(t)$, that determines whether non-adiabatic transitions will occur. We thus need to compare $\partial\vartheta/\partial t$ with $\hat D(t)$. Denoting the two time-dependent eigenvalues by $E_\pm(t)$, 
\[
E_\pm(t)=\frac{1}{2}(E_a+E_b)\pm\frac{1}{2}\sqrt{(E_a-E_b)^2+4|W_{ab}|^2}\,,
\]
the condition for adiabaticity becomes
\begin{equation}\label{eq:cond-adia}
\frac{1}{2}\left|\frac{\partial\vartheta}{\partial t}\right| \ll
\left|E_+(t)-E_-(t)\right|    \,.
\end{equation}
Expressing the Hamiltonian~\eqref{eq:HTLS} in terms of time-dependent Rabi frequency and detuning, 
\[
\hat H(t) = -\frac{1}{2}\left(\Delta_L(t)\hat\sigma_z
+\Omega_0(t)\hat\sigma_x\right)\,,
\]
we find 
\begin{equation}
  \label{eq:dot_theta}
\dot{\vartheta}=\frac{\Delta_L\dot\Omega_0 -\Omega_0\dot\Delta_L}{\Omega(t)}  
\end{equation}
with $\Omega(t)=\sqrt{\Omega_0^2(t)+\Delta_L^2(t)}$ the generalized Rabi frequency. In order to fulfill condition~\eqref{eq:cond-adia}, we need
\[
\frac{1}{2}\left| \Delta_L\dot\Omega_0 -\Omega_0\dot\Delta_L\right| \ll
\left(\Omega_0^2(t)+\Delta_L^2(t)\right)^{3/2}\,,
\]
which translates into the anticipated requirements of long and smooth pulses (small $\dot\Omega_0$, small $\dot\Delta_L$) and strong fields (large $\Omega_0$). While large detunings $\Delta_L$ are also good for adiabaticity, they are typically not helpful for the desired change in the instantaneous eigenstates. 

\begin{figure}[tbp]
    \centering
    \includegraphics[width=0.75\linewidth]{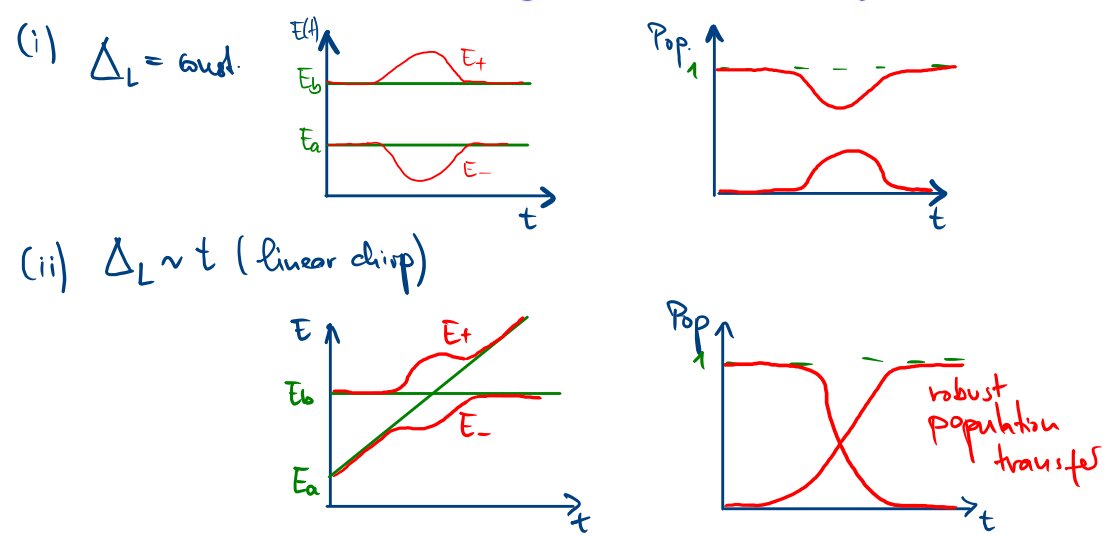}
    \caption{\textbf{Adiabatic following in a two-level system}, as reflected in the instantaneous eigenenergies $E_\pm(t)$ and the time-dependent population of the bare, or field-free, states, i.e., the eigenstates of $H_0$, $\ket{\phi_{a/b}}$.}
    \label{fig:adia}
\end{figure}
Two choices of control parameters to realize adiabatic following in a two-level system are sketched in Fig.~\ref{fig:adia}. In case (i), the detuning is kept fixed, and a pulse is applied, with its shape reflected in the time-dependence of the eigenvalues $E_\pm(t)$. In this case, if the system starts out in one of the eigenstates of $\hat H_0$, it will return to the same field-free state. A different result is obtained for a linear time-dependence of the detuning, as in case (ii), where, while adiabatically following one of the dressed states, $\ket{\phi_\pm(t)}$, the population is completely transferred between the bare states, $\ket{\phi_{a}}$ and $\ket{\phi_{b}}$. The latter is the famous Landau-Zener model, here realized with a linearly chirped pulse driving a two-level atom. Figure~\ref{fig:adia} highlights an advantage of adiabatic following -- it achieves \textit{robust} population transfer, irrespective of the exact shape or duration of the pulse. This is in contrast to control via a $\pi$-pulse which requires very precise pulse amplitude and duration. 

A popular and widely used example of adiabatic following is the so-called STIRAP  protocol, where STIRAP stands for ``stimulated Raman adiabatic passage''~\cite{VitanovRMP2017}. It realizes population transfer in a three-level system (see also App.~\ref{subsec:frameNLS}), where $\ket{1}$ is connected to $\ket{3}$ only via $\ket{2}$ which is, however, subject to fast decay. If one would apply the first (pump) pulse to drive the population from $\ket{1}$ to $\ket{2}$ and then the second (so-called Stokes) pulse to drive the transition from $\ket{2}$ to $\ket{3}$, the transfer could never reach a fidelity of one due to population decay when in $\ket{2}$. It turns out, however, that, depending on the detunings of the two pulses, one eigenvalue of the time-dependent Hamiltonian may vanish. The corresponding time-dependent eigenstate has non-zero projections only with the bare states $\ket{1}$ and $\ket{3}$~\cite{VitanovRMP2017}. If one can adiabatically follow this ``dark'' state, the decaying (and thus ``bright'') state $\ket{2}$ never gets populated, allowing for perfect population transfer from $\ket{1}$ to $\ket{3}$.
This requires both pulses to overlap in time and, counterintuitively, the Stokes pulse to precede the pump pulse~\cite{VitanovRMP2017}. It also requires strong fields to ensure adiabaticity.

Since large field strengths are not always available, let us inspect the alternative to ensure adiabaticity, counterdiabatic driving. The idea of counterdiabatic driving consists in adding an additional control $\hat H_{CD}(t)$ to the Hamiltonian $\hat H(t)$ that suppresses non-adiabatic transitions~\cite{DemirplakJPCA2003}. In order to calculate the counterdiabatic term for a two-level system, it is useful to start in a rotating frame in which $\hat H_{RWA}(t)=-\frac{\hbar}{2}\left(\Omega_0(t)\sigma_x-\Delta_L(t)\sigma_z\right)$ with $\Delta_L(t)=\omega_0-\omega_L(t)$, see App.~\ref{subsec:frameTLS}. With the counterdiabatic term, the Hamiltonian becomes $\hat H_{tot}(t)=\hat H_{RWA}(t)+\hat H_{CD}(t)$.
Transforming into the instantaneous eigenbasis of $\hat H_{RWA}(t)$, we obtain for the total Hamiltonian 
\[
  \hat H_{tot}^\prime(t) =
  \hat D(t) - \hat U(t)\frac{\partial \hat U^+(t)}{\partial t}
  + \hat U(t)\hat H_{CD}(t)\hat U^+(t)\,.
\]
Now $\hat H_{CD}(t)$ is chosen such that the last two terms cancel, that is, 
\begin{equation}
  \label{eq:H_CD}
  \hat H_{CD}(t) = i \frac{\partial \hat U^+(t)}{\partial t}\hat U(t)\,.
\end{equation}
For the example of the two-level system, we can use Eq.~\eqref{eq:UTLS} to evaluate the condition~\eqref{eq:H_CD} and obtain
\[
  \hat H_{CD}(t) = \frac{\dot\vartheta}{2}
  \begin{pmatrix}
    0 & i \\ -i & 0
  \end{pmatrix}=\frac{\dot\vartheta}{2}\hat\sigma_y\,.
\]
Recalling $\dot\vartheta$ given in Eq.~\eqref{eq:dot_theta}, the counterdiabatic drive has a Lorenztian shape and, being given by $\hat\sigma_y$ 
compared to $\hat\sigma_x$ in $\hat H_{RWA}$, it is phase-shifted compared to the original control. Transforming from the rotating frame back into the lab frame, this implies the counterdiabatic drive to oscillate as $\sin(\omega_L(t)t)$, when the original drive oscillates as $\cos(\omega_L(t)t)$~\cite{DemirplakJPCA2003}.
When calculating the counterdiabatic term for the STIRAP-Hamiltonian, ironically, the counterdiabatic term is found to drive the (non-existent) $\ket{1}\to\ket{3}$ transition~\cite{DemirplakJPCA2003}. This illustrates one of the limitations of counterdiabatic driving, a second limitation being that one needs to be able to diagonalize $\hat H(t)$ and determine $\hat U(t)$ in order to derive the counterdiabatic control term. 

\subsection{Optimal control theory: Fundamentals and state-to-state optimization}
\label{subsec:oct}
We now return to formal control theory~\cite{PontryaginBook} which is, in essence, an application of variational calculus to the dynamics of a quantum system. The starting point is the ability to solve the equations of motion for $\ket{\psi(t)}$, respectively $\hat\rho(t)$, analytically~\cite{BoscainPRXQ2021} or numerically~\cite{GlaserEPJD15,KochEPJQT22} or on a quantum device~\cite{MagannPRXQ2021}. In other words, optimal control theory constructs solutions of the control problem from the knowledge, even if implicit, of the system dynamics. In addition to the equations of motion, the second ingredient is the optimization functional, also referred to as cost functional or target functional. It formalizes the control target in terms of the figure of merit as well as additional constraints, for example on power consumption or smoothness of the controls. Indeed, all information on the physics of the control problem is encoded in these two ingredients --- optimization functional and equations of motion --- to an otherwise mathematical recipe. It is thus important to design the optimization functional as carefully as one constructs the equations of motion. 

Before discussing a first illustrative example, let us consider the general requirements that an optimization functional must meet~\cite{KochJPCM16,BasilewitschAQT19}. First of all, the optimization functional must be real-valued, such that it establishes an order relation that we can use to quantify improvement. It is customary, though not necessary, to normalize the functional to take values between 0 and 1. Second, it is useful to distinguish between so-called final-time costs and intermediate-time costs,
\begin{equation}
  \label{eq:functional}
  J[u(t)] = J_{t_f}[u(t)] + \frac{w}{t_f}\int_0^{t_f} g(\psi(t),u(t),t) dt\,.
\end{equation}
Typically, the final-time cost $J_{t_f}$ consists of the figure of merit to be optimized, whereas intermediate-time costs $g$ capture constraints on the dynamics. In Eq.~\eqref{eq:functional}, we have assumed that we consider the equation of motion for $\ket{\psi(t)}$, but generalization to $\hat\rho(t)$ is straightforward, 
$w\in \mathbb{R}$ is a weight, and the external controls are denoted by $u(t)$. $J[u(t)]$ is a functional of $u(t)$ in the sense of variational calculus, i.e., asking for an extremum of $J$, $\delta J =0$, allows us to determine the function (or ``shape'' of) $u(t)$. The third requirement that a well-defined functional $J[u(t)]$ must meet is that it takes its optimum value if and only if the optimum is reached. When writing $J[u(t)]$ as in Eq.~\eqref{eq:functional}, this implies that the figure of merit takes its minimum or maximum, depending on the optimization problem, and the second term vanishes, indicating that the constraints are enforced.

The simplest example for a control problem is to drive the system from a given initial state $\ket{\psi_0}$ to a desired target state $\ket{\psi_{target}}$. Assuming that the target state shall be reached at time $t_f$, this can be expressed as
\begin{equation}
  \label{eq:s-to-s-functional}
  J_{t_f}[u(t)]=\left|\bra{\psi_0}\hat U^+(t_f,0;u(t))\ket{\psi_{target}}\right|^2\,.
\end{equation}
Equation~\eqref{eq:s-to-s-functional} is very intuitive since $\hat U(t_f,0;u(t))\ket{\psi_0}$ is nothing but the system state at the final time that is obtained from time-evolving the initial state $\ket{\psi_0}$ under the external control $u(t)$, whatever $u(t)$ may be. Thus, $J_{t_f}$ measures the overlap of the actually obtained state $\ket{\psi(t_f)}$ with the desired state $\ket{\psi_{target}}$. In Eq.~\eqref{eq:s-to-s-functional}, we have taken the square modulus to ensure the real-valuedness of the functional; we could have equally well taken its real (or imaginary) part~\cite{PalaoPRA03}. 

A recent application has used Eq.~\eqref{eq:s-to-s-functional} to improve the preparation of circular Rydberg states in rubidium atoms~\cite{PatschPRA18,LarrouyPRX20}. These are states of the valence electron with a large value of the principal quantum number $n$ and the maximum value that the projection quantum number can take, $m_\ell=\pm(n-1)$. Here, we have made use of the close analogy between the hydrogen atom, where the state of the electron is described by the three quantum numbers $n, \ell, m_\ell$~\cite{Cohen-TannoudjiQM2}, and the valence electron in alkali atoms with the same set of quantum numbers. The name ``circular'' derives from the shape of the electron's orbit in that state; the interest in these states is due to their intrinsic protection against decoherence which makes them good candidates for quantum sensing~\cite{FaconNature2016} or quantum computing~\cite{XiaPRA2013,CohenPRXQ2021}. The protection arises from the very small number of electric dipole-allowed transitions that involve a circular state. This means, however, that it is also not easy to prepare a circular state. The challenge is in the large angular momentum transfer that is required to prepare a circular state.

\begin{figure}[tbp]
    \centering
    \includegraphics[width=0.95\linewidth]{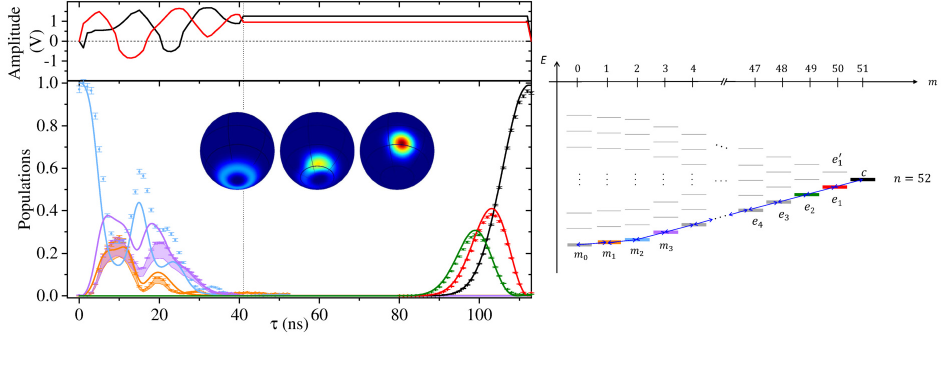}
    \caption{\textbf{Preparation of circular Rydberg state} with RF pulse  derived with optimal control theory: The pulse (with amplitudes of the two quadratures shown in the top left panel) transfers the population initially in $m_\ell=2$ to $m_\ell=51$. The lower left panel compares the experimentally measured populations (crosses) to the theoretical prediction (solid lines), with the colors encoding the $m_\ell$ states as shown on the right. The inset displays the Husimi-$Q$ functions of the initial state and the state after 15$\,$ns, respectively  $\approx 40\,$ns. The latter, also marked by the vertical line in the left panels, is the time by which a spin-coherent state has been prepared.  Adapted from Ref.~\cite{LarrouyPRX20}.}
    \label{fig:Rydberg}
  \end{figure}
Typically, one starts with laser excitation from the electronic ground state. Since the Rydberg state is accessed via a two-photon absorption, it results in a state with the desired value of the principal quantum number $n$ and $m_l=2$. One can then apply a radio-frequency (RF) field which quasi-resonantly drives a sequence of transitions between levels $m_l$ and $m_l+1$ until the circular state is reached~\cite{SignolesPRL2017}. This protocol can be rationalized in terms of a $\pi$-pulse, generalized from a two-level system to a system with $N=n-3$ levels, and visualized very similarly to the Bloch sphere representation of a two-level system, using the Husimi-$Q$ function~\cite{AgarwalPRA1998} for the $N$-level system. The circular state corresponds to the North pole of the generalized Bloch sphere. All the intermediate levels are not protected against decoherence which is why the generalized $\pi$-pulse should be as short as possible. It can, however, not be made arbitrarily short simply by increasing the pulse amplitude, unlike in the case of the two-level system: When the pulse amplitude becomes too large, the pulse drives off-resonantly also the transition from $m_\ell=2$ to $m_\ell=1$. When balancing the requirements of avoiding decoherence in the intermediate levels and minimizing population ``loss'' to $m_\ell=1$ in the best possible way, the circular state is prepared with a fidelity of about 80\%~\cite{LarrouyPRX20}.

With RF fields shaped according to the prediction from optimal control theory, the fidelity measured in the experiment, carried out by the cavity-QED group at Collège de France in Paris, increases to about 97\%~\cite{LarrouyPRX20}. This number is mainly limited by the accuracy with which the circular state was detected. Theoretically, the fidelity can be brought much closer to one. Knowing of the limited detection accuracy in the experiment, the calculations were stopped at 99\%~\cite{PatschPRA18}. Analysis of the optimized pulse revealed a two-stage protocol~\cite{PatschPRA18}: In the first 40$\,$ns, the RF field creates a spin-coherent state which subsequently is rotated to the North pole of the Bloch sphere in the second stage. The shape of the RF field reflects the two stages -- it consists of modulations during the first stage, whereas  constant amplitude is sufficient during the second stage. Note that a constant pulse  amplitude will typically not result from numerical optimization unless specifically enforced. In our case, we simply flattened the pulse based on the understanding of the dynamics in order to reduce pulse complexity and ease experimental implementation of the pulse~\cite{LarrouyPRX20}.

Encouraged by the seamless interplay of theory and experiment in the circular state preparation, we considered the preparation of a non-classical superposition state
$(\ket{m_\ell=1}+\ket{m_\ell=51})/\sqrt{2}$ as another example. For this control task,  no protocol was previously known. Similarly to the circular state preparation, a comparatively simple pulse shape was derived with optimal control theory~\cite{LarrouyPRX20}. It consists of modulations in the first 50$\,$ns and a linear chirp in the remaining part and results in an equally high fidelity of about 97\%, again limited by the experimental detection capability~\cite{LarrouyPRX20}.

One may wonder whether such a seamless interplay of theory and experiment was to be expected. The answer is yes! There are two reasons for this optimism: First of all, the Hamiltonian describing the valence electron in alkali atoms, that is, the energy levels but also all transition matrix elements, is known to high precision. In other words, our theoretical model of the system is very reliable. Second, the numerical optimization was carried out in a rotating frame. This eliminates all irrelevant timescales from the calculations and helps to avoid numerical artefacts, i.e., physically irrelevant features, in the pulse shape, reducing the overall pulse complexity.

But how does the optimization actually proceed? Let us return to the formal problem of determining the optimum of $J[u(t)]$. There exist two different approaches to solving the optimization problem numerically. Their key difference is whether they are based on $J$ alone or whether they also make use of the gradient of $J$ to construct changes to the controls $u(t)$~\cite{KochEPJQT22}. In the gradient-free approach, the controls $u(t)$ are parametrized, for example by expanding them into Fourier components (or any other suitable basis) and treating the frequencies and expansion coefficients as optimization parameters. The actual optimization is then carried out via a non-linear search in the parameter space, with suitable methods found for example in the NLopt software package~\cite{NLopt}. Gradient-free methods come with the advantage that the search is global in the parameter space. However, the search tends to get stuck when the number of parameters is large. In contrast, gradient-based approaches provide fast convergence, almost irrespective of the representation of the pulse. However they 
are local, performing the search with the information provided by the gradient around the initial ``guess'' pulse. The best of both worlds can be leveraged in hybrid optimization~\cite{GoerzEPJQT15}, consisting of a pre-optimization step using a gradient-free method with just a few parameters and a gradient-based optimization that uses the outcome of the pre-optimization as initial ``guess''.

The state preparation in Rydberg atoms highlighted above was carried out with a particular version of the gradient-based approach, Krotov's method~\cite{KrotovBook}. The quantum version of Krotov's method~\cite{ReichJCP12} has been our workhorse for quantum optimal control. Depending on the equations of motion, the figure of merit and additional constraints, Krotov's method provides us with a constructive way to design  an optimization algorithm, i.e., to derive a set of equations that are then solved numerically in an iterative way. The key advantage of Krotov's method is its guaranteed monotonic convergence, irrespective of the physics of the control problem. Krotov's method can be applied, in particular, to non-linearities, both with respect to the state in the equations of motion, including dissipation, and with respect to the controls; it also allows for non-convex target functionals. Having this powerful mathematical tool at hand, we are free to focus on the physics of the control problem, that is, on using the most appropriate equations of motion and finding the most suitable control target.

Before providing examples for control targets beyond preparing a certain desired state, let us look at one more level of detail as to what it entails to solve a quantum control problem by numerical optimization with Krotov's method~\cite{GoerzSciPost19}. 
As with any gradient-based method, ensuring an extremum of the optimization functional requires both $\delta_{u} J=0$ and $\delta_{\psi} J=0$ since $J$ depends both explicitly and implicitly, via the states, on the controls $u(t)$. The condition  $\delta_{\psi} J=0$ yields a dynamical equation for an auxiliary state, typically referred to as adjoint state, where the ``initial'' condition is given at the final time $t_f$. When the optimization targets a desired state, the initial condition is simply this state, $\ket{\psi_{target}}$; in general, it is derived from the figure of merit $J_{t_f}$. Since the initial condition is given at the final time, the adjoint state needs to be propagated backwards in time. The condition $\delta_{u} J=0$ yields the update rule for the external field. For Krotov's method where we minimize, as additional constraint, the change in the field, the rule is given by
\begin{equation}
  \label{eq:update}
  \Delta u(t)=u^{new}(t)-u^{old}(t) \sim \mathfrak{Im}\left[\bra{\psi_{target}}
    \hat U^+\left(t_f,t;u^{old}\right)\frac{\partial \hat H}{\partial u}
    \hat U\left(t,0;u^{new}\right)\ket{\psi_0}
    \right]\,.
\end{equation}
Here, $\hat U\left(t,0;u^{new}\right)\ket{\psi_0}$ implies that the initial state is forward propagated from time $0$ to $t$ using the new (updated) control $u^{new}$, whereas $\ket{\psi_{target}}$ is backward propagated from time $t_f$ to time $t$ using the old control $u^{old}$. The derivative $\partial \hat H/\partial u$ yields the operator that couples to the external field, typically the dipole moment, cf. Eqs.~\eqref{eq:H} or \eqref{eq:Hqubit}. Thus, Eq.~\eqref{eq:update} implies that, 
at time $t$, the dipole moment matches the forward and backward propagated states to yield the change in the control $\Delta u(t)$ at that instant of time. The requirement to both forward and backward propagate is a hallmark of any gradient-based optimization in quantum control. The main difference between gradient-based algorithms is the way in which the update is applied --- for Krotov's method, sequentially for each instant in time, as in Eq.~\eqref{eq:update}, whereas for GRAPE~\cite{KhanejaJMR05}, another popular algorithm, the update is applied concurrently for all $t$ at once. At  first glance,  Eq.~\eqref{eq:update} appears to be implicit in $u^{new}(t)$, but it is sufficient to use two different time grids, shifted by half the numerical time step, for the states and the control~\cite{PalaoPRA03}, in order to solve  Eq.~\eqref{eq:update}. Use of optimal control theory is thus equivalent to iteratively solving Eq.~\eqref{eq:update}, or generalizations thereof, starting from a first ``guess'' for  $u^{old}(t)$. 

\subsection{Optimal control theory: Optimization functionals}
\label{subsec:functionals}

How does Eq.~\eqref{eq:update} change when we use a different functional $J$ in  Eq.~\eqref{eq:functional}? For example, a relevant application in the context of quantum technologies is the implementation of a desired unitary evolution~\cite{PalaoPRA03} such as a gate in quantum computation. At first glance, one may think of it as simultaneous state-to-state transformations, all carried out by the same control $u(t)$. The number of states is given by the size of the Hilbert space on which the desired evolution is defined, i.e., $N=2$ for a single-qubit gate or $N=4$ for a two-qubit gate. However, the proper implementation of the desired unitary requires more than $N$ simultaneous state-to-state transitions --- also the relevant phase relations between the states must be ensured~\cite{PalaoPRA03}. This is achieved by generalizing the scalar product between the desired target state $\ket{\psi_{target}}$ and the propagated initial state $\hat U\left(t_f,0;u(t)\right)\ket{\psi_0}$ in Eq.~\eqref{eq:s-to-s-functional} to the Hilbert-Schmidt product for operators. Denoting the desired unitary evolution or quantum gate by $\hat O$, the final-time cost thus becomes
\[
  J_{t_f}=1-\frac{1}{N}\mathfrak{Re} \left[\tr\left\{\hat O^+ \hat U\left(t_f,0;u(t)\right)\right\}\right]\,,
\]
where we assume, without loss of generality, a minimization problem, i.e., the optimal value that $J$ can take is zero. Often, qubits are carried by physical objects such as atoms or superconducting qubits with a size of Hilbert space much larger than $N=2$. We then need to project the actually implemented evolution onto the subspace on which $\hat O$ is defined, 
\begin{equation}
  \label{eq:Jgate}
  J_{t_f}=1-\frac{1}{N}\mathfrak{Re} \left[\tr\left\{\hat O^+ \hat P_N \hat U\left(t_f,0;u(t)\right)\hat P_N\right\}\right]\,.
\end{equation}
Expanding out the trace in Eq.~\eqref{eq:Jgate}, we see the similarity  to Eq.~\eqref{eq:s-to-s-functional}. The generalization of Eq.~\eqref{eq:update} is then also straightforward,
\[
  \Delta u(t)\sim \mathfrak{Im}\left[\frac{1}{N}\sum_{k=1}^N\bra{\psi_{k}}\hat O^+
    \hat U^+\left(t_f,t;u^{old}\right)\frac{\partial \hat H}{\partial u}
    \hat U\left(t,0;u^{new}\right)\ket{\psi_k}
    \right]\,.
\]
Here, the backward propagation concerns the target states $\hat O\ket{\psi_k}$, and 
the update $\Delta u(t)$ is given by summing over all $N$ states. In other words, the optimization of a quantum gate requires the forward, respectively backward, propagation of $2N$ states instead of just two. For a small number of qubits, this 
can easily be done on classical computers, and various software is available, for example GRAPE within QuTiP~\cite{qutip,qutip2} or Krotov's method as separate package~\cite{GoerzSciPost19}, whereas for larger systems, a hybrid quantum-classical approach is necessary~\cite{MagannPRXQ2021}.

When the control task is to implement an entangling operation for two qubits, it may be advantageous to exploit the concept of local equivalence in the optimization -- two two-qubit gates $\hat A$ and $\hat B$ are said to be locally equivalent when applying single-qubit (``local'') gates before and after $\hat A$ results in $\hat B$ and vice versa. For example, the controlled phasegate and the controlled NOT gate are locally equivalent. Assume that you want to implement a CNOT gate but the Hamiltonian governing the dynamics of the qubits is diagonal in the logical basis (as is the case, for example, for Rydberg atoms). An optimization targeting a CNOT gate will then fail, whereas an optimization targeting the local equivalence class of CNOT will be successful~\cite{MuellerPRA11}. The corresponding target functional is expressed in terms of the so-called local invariants, three real numbers that characterize each local equivalence class, more precisely, in terms of the Euclidean distance between the local invariants of the desired and the actually realized gate~\cite{MuellerPRA11}. 

\begin{figure}[tbp]
  \centering
  \includegraphics[width=0.75\linewidth]{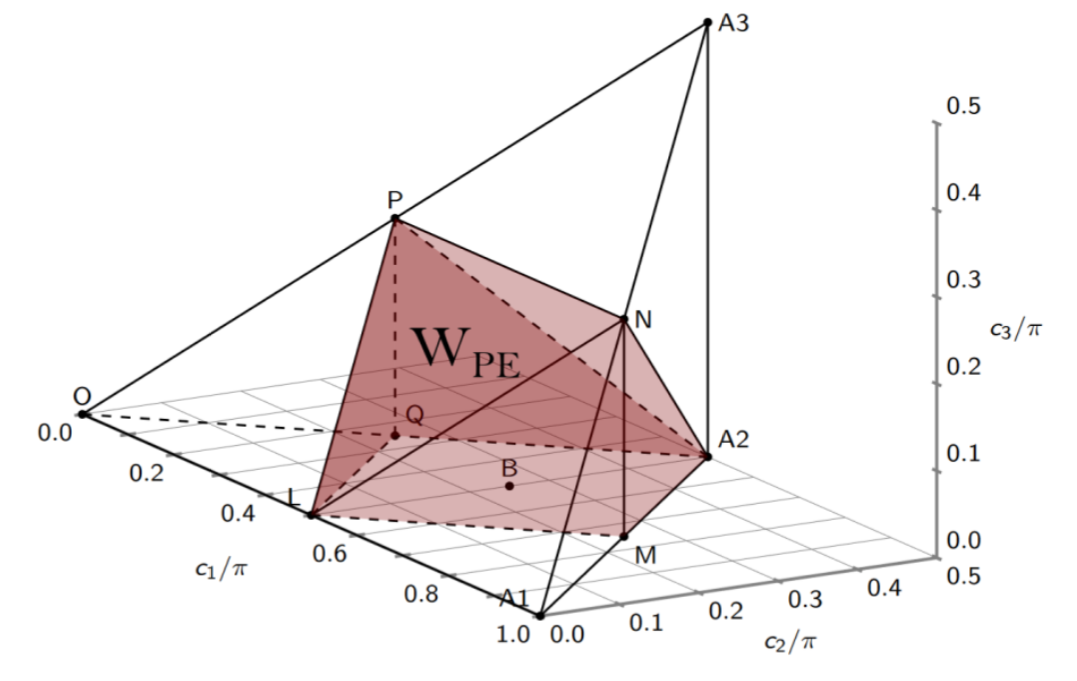}
  \caption{\textbf{The polyhedron of perfect entanglers}, $W_{PE}$, comprising all two-qubit gates $\hat U=\hat k_1e^{i(c_1\sigma_x\sigma_x+c_2\sigma_y\sigma_y+c_3\sigma_z\sigma_z)} \hat k_2$ with the property of transforming at least one separable state into a maximally entangled state, can be used as an optimization target~\cite{WattsPRA15}. Here, $\hat k_{1/2}$ are single-qubit operations, and the non-local content is completely characterized by the Weyl chamber coordinates $c_1,c_2,c_3$, with the point $O$ corresponding to the identity, and $L$ to the CNOT gate. Optimal control targeting a perfect entangler minimizes the distance to $W_{PE}$~\cite{GoerzPRA15}.}
  \label{fig:weylchamber}
\end{figure}
The idea of targeting a whole class of quantum gates instead of a single specific one can be generalized further. This is interesting in the context of universal quantum computation where any $n$-qubit unitary can be realized by a sequence of single qubit gates and one entangling two-qubit gate~\cite{NielsenChuang}. These gates form a finite set, termed universal set of gates~\cite{NielsenChuang}. Every entangling operation requires an interaction between the qubits; but given an interaction, it is often not immediately obvious which entangling operation is most straightforwardly implemented. This question can be answered using optimal control targeting 
an arbitrary perfect entangler. Here, the target functional minimizes the distance of the local invariants of the realized gate to the space of perfect entanglers~\cite{WattsPRA15,GoerzPRA15}, shown as the shaded area in Fig.~\ref{fig:weylchamber}. The result of the optimization is then not only the control $u(t)$ that implements an entangling gate but also the gate itself. 

These control targets --- a specific two-qubit gate, a local equivalence class of gates or a two-qubit gate that is a perfect entangler --- are expressed by functionals of increasing complexity. Their advantage is that we state the control problem in a more general way which may ease finding a solution. In principle, any physical property can be turned into an optimization functional. For example, one can minimize energy~\cite{DoriaPRL11} or maximize entropy~\cite{BasilewitschNJP17} or squeezing~\cite{HalaskiPRA24}, instead of asking to prepare a specific state with minimal energy or maximal entropy or squeezing. This is advantageous as soon as more than one state with the optimal value of the desired property exists, since the optimization can explore all possible target states.

The desired property can be targeted at the final time only or at intermediate times as well. In other words, there is no fundamental difference between targets and constraints in optimal control theory, which is why in the mathematical literature, they are jointly referred to as ``costs''. 
However, targeting a property at several or all intermediate times is a much harder control problem than one with a \textit{single} final-time ``cost'' $J_{t_f}$~\footnote{Moreover, the numerical difficulty increases, as the equation of motion for the backward propagation contains a source term for time-dependent targets.}. One can rationalize this as follows: In order to solve a control problem, we should first make sure that the problem is solvable. The answer to this question --- yes or no --- is provided by controllability analysis to which we will get below in Sec.~\ref{subsec:controllability:basics}. In controllability analysis, one assumes to have unlimited resources at the disposal, for example unlimited time (as long as it is finite) or unlimited pulse energy. The result of controllability analysis will be negative if a symmetry precludes reaching the target; it thus reflects fundamental properties of the system. If the answer is positive, i.e., the quantum system is controllable, then very often many solutions to a given control problem exist --- for unlimited resources. As soon as one restricts the resources, the number of solutions will shrink~\cite{MooreJCP12}, possibly all the way to zero. In the case of the resource time, the transition point to no more solution is referred to as ``quantum speed limit'' or ``minimal evolution time'', and we will see examples of it below in Sec.~\ref{sec:applications}. Including constraints in the optimization functional is another way to limit the resources available for solving the control problem~\cite{MooreJCP12}. Conversely, stating the control problem in as general terms as possible will likely increase the search space and the resources. It is therefore important to design the optimization functional in a way that captures all of the desired physics while corresponding to the least constraints on the problem; it will help to make the problem solvable in practice.

\subsection{Controllability}
\label{subsec:controllability:basics}

Is a given control problem solvable, assuming that all necessary resources can be provided?
The mathematical tools for answering questions such as ``which  states can be prepared?'' or ``which operations (such as gates) can be implemented?'' for a given quantum system are summarized under the term controllability analysis~\cite{Dalessandro2008}. For closed quantum systems with unitary time  evolution, the answer to these questions is determined by the Hamiltonian~\eqref{eq:H} alone. In other words, we do not need to solve the equations of motion but can infer the answer  directly from $\hat H=\hat H_0+\hat H_I$~\footnote{The time-dependence of $\hat H(t)=\hat H_0+\hat H_I(t)$ is omitted here, since $\hat H_I(t)=\sum_j u_j(t)\hat H_j$ and what matters for controllability are the properties of the $\hat H_j$.}. More precisely, we need to inspect the nested commutators of $\hat H_0$ and $\hat H_I$, the drift and control Hamiltonians. Why this is sufficient, is readily understood for finite-dimensional systems by visualizing the system state in the (generalized) Bloch sphere~\cite{BertlmannJPA08}: Unitary evolutions are rotations of the state vector $\ket{\psi}$ on the sphere. These rotations are generated by the Hamiltonian. For infinitesimal evolutions $\hat U(t+\delta t,t)\approx \openone - i (\hat H_0+\hat H_I)\delta t$, with $\hat H_0$ and $\hat H_I$ being two vectors lying in the tangential plane attached to the Bloch sphere at the point corresponding to $\ket{\psi(t)}$. The terms in the Hamiltonian thus define the directions in which the state vector can move (they are indeed elements of a vector space). The ability to reach any point on the sphere then corresponds to the ability to generate motion in all possible directions. For finite evolution times, the directions are not just given by the drift and control Hamiltonians, but also by their (nested) commutators. This can easily be seen when expanding the exponent, $\hat U(t,0)=\exp\left(i (\hat H_0+\hat H_I) t\right)$~\footnote{More precisely, we should account for time ordering, $\hat U(t,0)=\mathcal T \exp\left(i \left(\hat H_0+\hat H_I(t)\right) t\right)$, but again, this does not affect the controllability of a system.}. The (nested) commutators are elements of the same vector space as $\hat H_0$ and $\hat H_I$. The question of controllability has thus been translated into the question of how many linearly independent vectors are generated by expanding the exponent.

In more formal terms, inspecting the nested commutators of $\hat H_0$ and $\hat H_I$ corresponds to constructing the dynamical Lie algebra of the quantum system~\footnote{In general, an algebra is a vector space with ``multiplication''; in case of a Lie algebra, the ``multiplication'' is given by the commutator.}. The Lie algebra is said to be of full rank if its dimension is equal to $N^2$ or $N^2-1$, where $N$ is the dimension of Hilbert space~\cite{Dalessandro2008}. The Lie algebra is then isomorphic to $\mathfrak{u}(N)$, respectively $\mathfrak{su}(N)$, the algebras associated to the unitary group $U(N)$ (containing all $N\times N$ unitary matrices), respectively the special unitary group $SU(N)$ (containing all $N\times N$ unitary matrices with determinant 1). The difference between the two cases concerns merely the physically irrelevant global phase associated with $\openone$. If the dynamical Lie algebra is of full rank, any element of the group $U(N)$, respectively $SU(N)$, can be realized, i.e., the quantum system is (completely, or evolution operator-) controllable~\cite{Dalessandro2008}.  

Since the dynamical Lie algebra is a vector space, one way to check the full rank-condition is to construct an orthogonal basis of that vector space from the commutators, thus determining its dimension. Alternatively, the Hamiltonian can be represented on a graph. The full rank condition can then be inferred from the connectedness of the graph~\cite{Dalessandro2008}.
%In either way, the exponential scaling of the Hilbert space dimension with number of particles presents a challenge to analysing controllability. 
While traditionally controllability analysis has been used for given, fixed Hamiltonians, we will see below in Sec.~\ref{subsec:controllability} that we can turn this perspective around and ask what is the minimal number of terms in $\hat H_I$ that yield a completely controllable system. This extends quantum control from the question of how to design the external fields $u(t)$ to that of how to overall engineer the field-matter coupling $\hat H_I(t)=\sum_j u_j(t)\hat H_j$.

The extension of controllability analysis to open quantum systems is extremely challenging and has largely been confined to systems with Markovian dynamics.  The semi-group structure of such dynamics is amenable to Lie algebraic tools which has allowed e.g. for characterizing reachable sets of states~\cite{ToSHRMP09}. 

\section{Selected applications}
\label{sec:applications}

Having laid out the basic concepts in Sec.~\ref{sec:basic}, we can now inspect some recent applications of quantum control. The selection is heavily biased towards work from my group but Refs.~\cite{GlaserEPJD15,KochEPJQT22} provide a more comprehensive overview. Indeed, optimal control theory, and quantum control more generally, have proven useful, in the past decades, for applications in quantum information science across all physical platforms from nuclear magnetic resonance to atomic physics or solid state qubits~\cite{GlaserEPJD15,KochEPJQT22}. A key motivation for using optimal control theory is that quantum devices are open quantum systems. While advanced engineering aims to isolate a device's desired quantum features, the need to execute operational and readout protocols implies an inevitable coupling of the device to its environment. Optimal control theory is a tool allowing us to identify the best possible balance between desired control and undesired disturbance, i.e.,  decoherence.

In addition to its promise to help make quantum technologies practical, 
the control of open quantum systems tackles questions that are also of fundamental interest and relevant beyond specific applications~\cite{KochJPCM16}. For example, what are viable control strategies in the presence of noise and how do they depend on the system-environment coupling? What are fundamental limits to quantum control in terms of attainable targets or errors? While general answers to these questions are still lacking, a set of preliminary useful rules has already been identified~\cite{KochJPCM16}:
\begin{enumerate}\label{list:rules}
\item If the desired operation shall keep pure states pure and the system dynamics are Markovian, the effect of the environment is detrimental. Then the best control strategy is to avoid decoherence. We will discuss examples below in Sec.~\ref{subsec:open:fight}.
\item If the desired operation changes the system purity, the control target can only be realized thanks to the presence of the environment. Then Markovian
dynamics may be desired --- the control target is reachable if it is a fixed point of the Liouvillian, and external controls can be used to ensure that this is the case. The corresponding control strategy is sometimes referred to as quantum reservoir engineering~\cite{PoyatosPRL96}. Section~\ref{subsec:open:exploit} will show how optimal control theory can be used in this setting.
\item If the realization of the control target requires dissipation which is not naturally available, we can create it by coupling the quantum system of interest to auxiliary degrees of freedom that are then simply discarded or measured. We will discuss this strategy in Sec.~\ref{subsec:open:create}.
\end{enumerate}
When the dynamics of the open quantum system is non-Markovian, the environment may have both beneficial and detrimental effects on control. Examples of beneficial effects are a larger number of implementable gate operations~\cite{ReichSciRep15} or faster operation~\cite{DeffnerReview2017}. By and large, however, control of systems with non-Markovian dynamics is uncharted territory and beyond the scope of this tutorial.

Before we can examine opportunities for and limitations to control in open quantum systems, we need to take one step back, however, and answer the question how we can quantify success of the control. In other words, what are suitable figures of merit for open quantum systems? While the target functionals discussed in Sec.~\ref{subsec:functionals} above provide a good starting point, there are also some  pitfalls to avoid.

\subsection{Control of open quantum systems: How to quantify success}
\label{subsec:open:figure-of-merit}

\begin{figure}[tbp]
  \centering
  \includegraphics[width=0.35\linewidth]{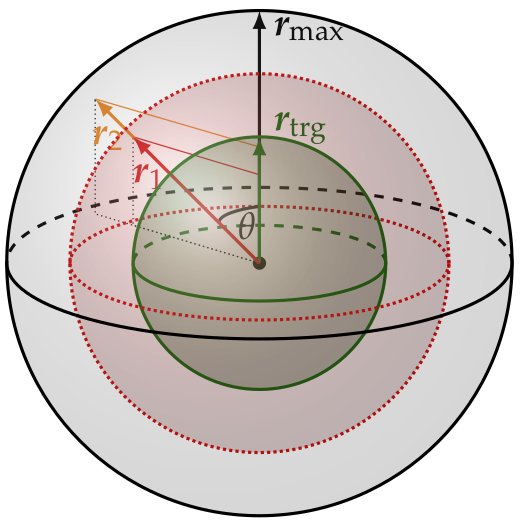}
  \caption{\textbf{Which state matches a desired target state better?}
    Bloch sphere representation of matching states $\hat\rho_{1/2}$ with Bloch vectors $\vec r_{1/2}$ to a \textit{mixed} target state with Bloch vector $\vec r_{trg}$. While a true distance measure quantifies $\vec r_1$ as closer to $\vec r_{trg}$ than $\vec r_2$, the Hilbert Schmidt product measures only the projection onto the axis of $\vec r_{trg}$ and thus is larger for $\vec r_2$. This failure of the Hilbert-Schmidt product for a mixed target state  can be remedied by a figure of merit that matches both angle and length of the target Bloch vector. 
    Adapted from Ref.~\cite{BasilewitschAQT19}.}
  \label{fig:mixed_bloch}
\end{figure}
When the target of the control is to reach a given desired state, we can check whether the state of the system after time evolution matches this state by taking the overlap and turning it into a real number, cf. Eq.~\eqref{eq:s-to-s-functional}. Moving to open quantum systems, Hilbert space kets become density operators, and naively, we could replace the scalar product between the time-evolved and the target state in Eq.~\eqref{eq:s-to-s-functional} by the Hilbert-Schmidt product for the respective density operators, i.e., $J_T\sim \tr\left\{\hat\rho(T)\,\hat\rho_{target}\right\}$. This works as long as the target $\hat\rho_{target}$ is a pure state but fails if $\hat\rho_{target}$ is mixed~\cite{BasilewitschAQT19} because the Hilbert-Schmidt product $\tr\left\{\hat\rho_1\,\hat\rho_2\right\}$ is not (unlike the Hilbert-Schmidt distance, $\frac{1}{2}\tr\left\{\left(\hat\rho_1-\hat\rho_2\right)^2\right\}$) a distance measure.

Visualizing the states on the Bloch sphere provides an intuitive understanding of the failure~\cite{BasilewitschAQT19}, cf. Fig.~\ref{fig:mixed_bloch}. It makes use of the fact that any state $\hat\rho$, defined on a finite-dimensional Hilbert space (of dimension $N$),  can be expressed in a complete orthonormal operator basis~\cite{BertlmannJPA08},
\[
  \hat\rho=\frac{\openone}{N}+\sum_kr_k\hat A_k
  =\frac{\openone}{N}+\vec r\cdot\vec{\hat A}\,.
\]
In terms of Bloch vectors, the Hilbert-Schmidt product, or overlap, of two states is given by
\[
\tr\left\{\hat\rho_1\,\hat\rho_{target}\right\}=\frac{1}{N}+|\vec r_1| |\vec r_{target}| \cos\theta\,,
\]
where $\theta$ is the angle between the Bloch vectors, see Fig.~\ref{fig:mixed_bloch}. Assuming $\vec r_1\parallel \vec r_2$ and $|\vec r_2| > |\vec r_1 | > |\vec r_{trg} |$ as shown in Fig.~\ref{fig:mixed_bloch},  the purer state $\vec r_2$ has a larger projection onto the target state and thus a larger Hilbert-Schmidt product, although $\hat \rho_2$ is ``farther'' away from the target than $\hat\rho_1$. Indeed, optimizing for $\hat\rho_{trg}$ with the Hilbert-Schmidt product will yield the pure state with $\theta=0$ instead of $\rho_{trg}$. This can be remedied by employing a true distance measure such as the Hilbert-Schmidt distance or the trace distance.
This comes, however, with a potential numerical instability due to the discontinuity of the square root. Alternatively, the intuitive picture of Fig.~\ref{fig:mixed_bloch} suggests to optimize for the desired length and angle of the target Bloch vector~\cite{BasilewitschAQT19}. This has been used in an application of optimal control theory to an optomechanic system where the mixed target state  balances purity and squeezing~\cite{BasilewitschAQT19}.

When targeting a quantum gate, the naive approach would be to lift Eq.~\eqref{eq:Jgate} from Hilbert space to Liouville space~\cite{ToSHJPB11}. 
The trace implies that a complete orthogonal basis of the logical Liouville space, $\{\hat\rho_j\}_{j=N^2}$, 
needs to be propagated and, after time evolution, compared with the target operation applied to that basis state, $\hat\rho_j(T)\hat O\hat\rho_j\hat O^+$. Here $N=2^n$ is the dimension of the (logical) Hilbert space for $n$ qubits (the physical Hilbert space of each qubit may comprise more than two levels). This approach involves significantly more effort than necessary because a complete basis of Liouville space needs to be used only if the optimization target is an arbitrary open system evolution. When instead the target is a quantum gate, we can exploit that the desired evolution is unitary, or as close to unitary as possible. In this case, it is sufficient to use three judiciously chosen initial states and compare their time evolution to the desired operation~\cite{ReichPRA13,ReichPRL13,GoerzNJP14}.

It might be very surprising that the number of states required to verify the desired operation is independent of the system size. This can be rationalized as follows. Assume for a moment that the time evolution \textit{is} unitary. In this case the question is: How many states are necessary to infer from their time evolution whether a desired unitary is implemented? Since all unitaries for a given system form a group, it is equivalent to asking: How many states are needed to distinguish any two unitaries? The answer is two~\cite{ReichPRA13}, and these two states have a clear geometric interpretation. When assessing whether two unitaries are equal, we need to determine whether they share an eigenbasis and whether their eigenvalues (or eigenphases) are equal. A single state built from one-dimensional orthogonal projectors $\hat P_i$, $\hat\rho_B=\sum_i\lambda_i\hat P_i$ with $\lambda_i\neq\lambda_j$ for all $i\neq j$, is sufficient to ``fix'' the basis. A second state, that is rotated with respect to all $\hat P_i$,  $\hat\rho_P=\hat P_{rot}$ with  $\hat P_{rot}\hat P_i\neq 0\;\forall i$, answers the question about the eigenphases. Note that the rotated state is one element of a basis that is mutually unbiased with respect to the basis of the $\hat P_i$. In order to generalize this to open system evolution, a third state is necessary to quantify the non-unitarity of the evolution. This quantification is achieved by measuring the unitality on the logical subspace, $\hat\rho_3=\openone/N$~\cite{ReichPRA13}. The construction of these three states is useful not just for the figure of merit in gate optimization but also for determining the fidelity of gates or channels of actual quantum devices~\cite{ReichPRL13}. In gate optimizations, the most efficient approach is typically to replace $\hat\rho_B$, which is mixed, by the $N$ projectors from which $\hat\rho_B$ is constructed~\cite{GoerzNJP14}. In this case, $\hat\rho_3$ is not needed since non-unitarity can be inferred from the (non-unitary) time evolution of the $N+1$ initially pure states. The better numerical performance is rationalized by a more uniform distribution of the relevant information over the propagated states~\cite{GoerzNJP14}. While the number of states in this variant scales with system size, there is still an exponential saving compared to the naive approach~\cite{ToSHJPB11}. The same exponential saving carries over to protocols for determining gate or channel fidelities~\cite{ReichPRL13}.

Besides targeting specific states or gates with quantum optimal control, it may also be useful to optimize towards more general properties. Examples were local equivalence classes of two-qubit gates or the set of all perfect entanglers in Sec.~\ref{subsec:functionals} above. In order to adapt the corresponding  optimization functionals to open system evolution, we have to reconstruct the unitary part of that evolution from knowledge of the propagated states. Similarly to gate optimization, this can be done using $\hat\rho_B$ and $\hat\rho_P$ or $N+1$ one-dimensional projectors as initial states, provided the evolution is not too noisy (in which case the question about the unitary part looses its meaning)~\cite{Romer25}. The reconstruction algorithm, just as the fidelity estimation discussed above, is useful not just in quantum optimal control theory but also to benchmark actual devices. 

Having adapted our figures of merit to open quantum systems, the optimization proceeds as in the closed system case, except for the fact that, instead of the time-dependent Schrödinger equation, the corresponding equation of motion for the open system must be solved (for both forward and backward propagations)~\cite{GoerzNJP14}. 
With the necessary tools for studying the control of open quantum systems established, we can now look at examples (taken from the work of my group) which illustrate the three rules identified at the beginning of Sec.~\ref{sec:applications}.

\subsection{Control of open quantum systems: Avoiding decoherence}
\label{subsec:open:fight}

Avoiding decoherence is the best control strategy when the system dynamics are
Markovian and the control target is reached without any change of entropy. This is true for coherent state transformations including the implementation of quantum gates. In general, there are two ways to avoid decohence --- one is to ``beat'' decoherence by operating fast, the other one is to utilize protection from symmetries.

\begin{figure}[tbp]
  \centering
  \includegraphics[width=0.9\linewidth]{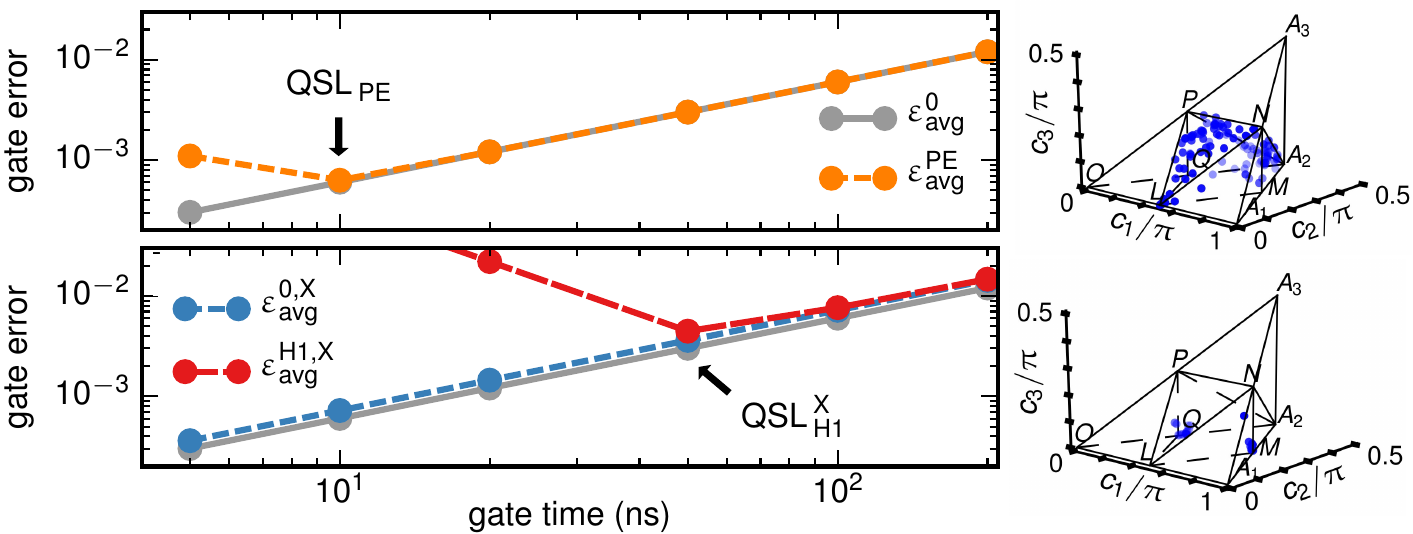}
  \caption{\textbf{Quantum speed limit for a universal set of gates for transmon qubits} (left): The gates are implemented via a shaped microwave drive, designed with quantum optimal control theory, that acts on the cavity to which both superconducting qubits are coupled. Using an optimization functional that targets an arbitrary perfect entangler, cf. Fig.~\ref{fig:weylchamber}, has allowed us to identify $\sqrt{iSWAP}$ as the ``natural'' gate for this architecture (right): For long gate times (200 ns in the upper panel), almost all perfect entanglers can be realized, whereas only the $\sqrt{iSWAP}$-gate survives when reducing the gate time (10 ns in the lower panel).
  Adapted from Ref.~\cite{GoerzNPJQI17}.}
  \label{fig:univ-set}
\end{figure}
We have already discussed examples of fast operations above, with the examples of Rydberg state preparations~\cite{PatschPRA18,LarrouyPRX20} in Sec.~\ref{subsec:oct}. Another example is the implementation of a universal set of gates, defined as the collection of single-qubit gates and one entangling gate from which any desired unitary can be constructed~\cite{NielsenChuang}.
Optimal control theory is ideally suited to determine the quantum speed limit~\cite{DoriaPRL11}
%for such a univeral set, i.e., the shortest possible duration with which all of these gates can be implemented.Since the speed limit
which can be quite different for different gates. It is therefore crucial to establish the quantum speed limit for the complete universal set, rather than relying on only one or two specific gates.
This is particularly true for superconducting qubits where it is \textit{a priori} not clear, due to residual couplings between the qubits,  whether the single-qubit gates or the two-qubit gate determine the speed limit. Moreover, a variety of entangling gates, each with a possibly different speed limit, can be implemented. In order to determine the fastest universal set for superconducting transmon qubits~\cite{GoerzNPJQI17}, 
we have used a combination of the tools introduced above, that is, optimization targeting an arbitrary perfect entangler~\cite{WattsPRA15,GoerzPRA15} and hybrid optimization~\cite{GoerzEPJQT15} combining a global search of the control landscape~\cite{NLopt} with Krotov's method~\cite{KrotovBook,ReichJCP12,GoerzSciPost19}.
To minimize the impact of external control noise, all gates are implemented via microwave drives that act on the common cavity mode coupling both transmons in the transmission line. For this architecture, manually designed gate protocols typically rely on avoiding cavity excitation which allows the cavity to be eliminated to derive an effective two-qubit Hamiltonian. However, operating in this dispersive limit entails a trade-off --- the weak coupling results in slower gate operation. A numerical approach to gate design using quantum optimal control has allowed us to forgo the dispersive limit~\cite{GoerzNPJQI17}. Exploring the full landscape of design parameters, specifically couplings, detunings and anharmonicities, we have determined the ``quasi-dispersive straddling qutrits'' regime to be optimal for the rapid creation and removal of entanglement necessary for a universal set of gates~\cite{GoerzNPJQI17}. Optimizing for an arbitrary perfect entangler instead of a specific two-qubit gate has furthermore allowed us to determine the ``natural'' gate of the architecture ---  only the $\sqrt{iSWAP}$ can be realized within 10$\,$ns, whereas for gate durations as long as 200$\,$ns almost any perfect entangler can be implemented~\cite{GoerzNPJQI17}, cf. Fig.~\ref{fig:univ-set} (right). We have also found that single-qubit gates require more than 10$\,$ns, putting the  quantum speed limit for the universal set at 50$\,$ns, cf. Fig.~\ref{fig:univ-set} (left). For such a gate duration, several entangling gates are possible, and we have opted for the B-gate as the perfect entangler that allows for the shortest circuits when decomposing an arbitrary unitary into the gates of universal set~\cite{ZhangPRL04}. Finally, while operating at the quantum speed limit is optimal for  minimizing decoherence, it comes at the expense of spectrally broad and temporally complex pulse shapes which may be hard to benchmark on an actual device. Extending the gate durations may also be beneficial in view of better robustness of the protocols with respect to parameter uncertainties.
These points highlight the complexity of optimizing the operation of quantum devices where a number of desiderata need to be balanced.

\begin{figure}[tbp]
  \centering
  \includegraphics[width=0.75\linewidth]{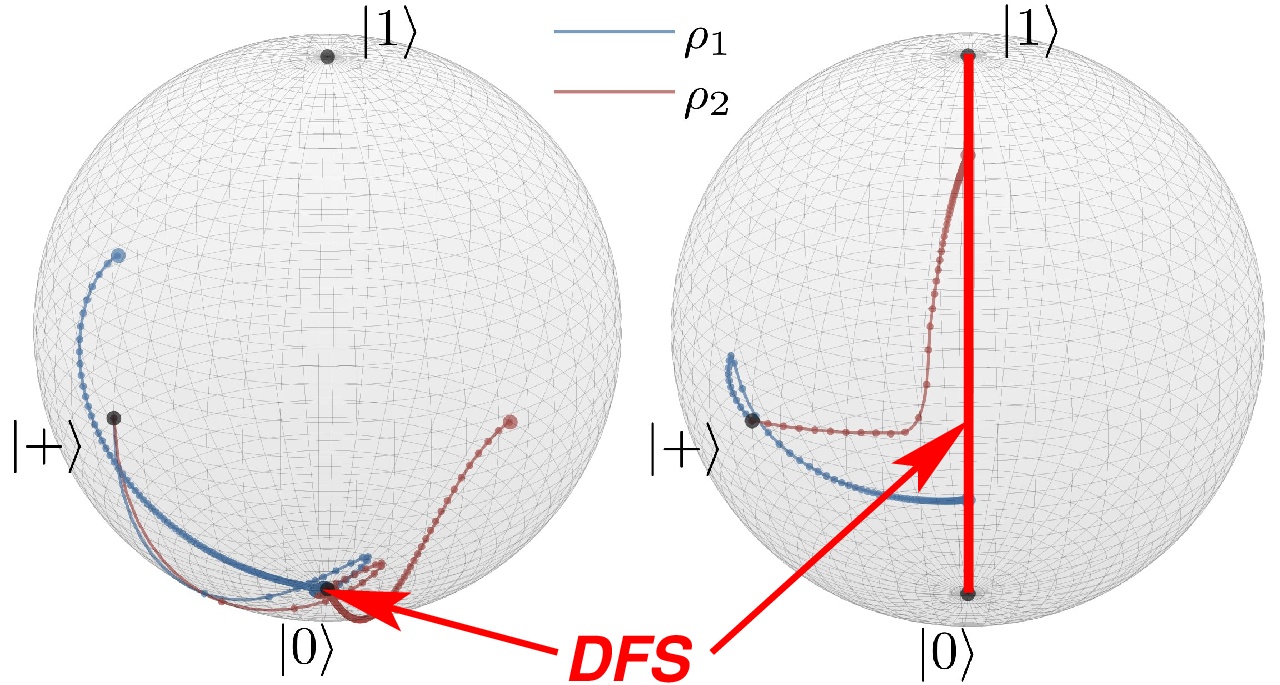}
  \caption{\textbf{Keeping the dynamics decoherence-free:}
    Time evolution of two states ($\rho_1$, $\rho_2$) that evolve under slightly different Hamiltonians, driven by the same $\sigma_x$-drive that was optimized to maximize the final-time distinguishability. The dynamics stays close to a decoherence-free subspace, the state $\ket{0}$ for $T_1$ decay (left) and the $z$-axis for $T_2$ dephasing (right), for a significant part of the protocol. Adapted from Ref.~\cite{BasilewitschPRR20}.}
  \label{fig:sensing-dfs}
\end{figure}
Besides operating fast, we can avoid decoherence by utilizing protection from symmetries, that is, by keeping the dynamics in decoherence-free subspaces~\cite{LidarDFS03}. With the time-dependent ``dark state'' in STIRAP, we have already encountered a one-dimensional example of a decoherence-free subspace~\cite{LidarDFS03} in Sec.~\ref{subsec:adiabatic} above.
But even though there are several equivalent ways to identify decoherence-free subspaces, being able to do this in practice may be hampered by system complexity. Here is where quantum optimal control can help. Optimizing directly for a decoherence-free subspace results in optimization landscapes with a very large number of local minima. In contrast, targeting the desired operation while accounting for decoherence via the equation of motion is typically sufficient to find decoherence-free subspaces, even approximate ones. For example, we have found that pulses designed to maximize the distinguishability of two quantum states that evolve under slightly different Hamiltonians, a prototypical question in quantum sensing,  steer the states close to a decoherence-free subspace~\cite{BasilewitschPRR20,QinSciPostPhys22}. The optimized pulses yield a distinguishability that is significantly larger than that obtained with a standard Ramsey protocol. The improvement translates also into a metrological gain in terms of the quantum Fisher information. 
Remarkably, the distinguishability of the two states can be stabilized to protocol durations that are orders of magnitude larger than the timescale of decoherence, for both $T_1$ decay and $T_2$ dephasing~\cite{BasilewitschPRR20}. 
While surprising at first glance, we could  rationalize this finding in terms of staying close to a decoherence-free subspace (the ground state for  $T_1$ decay and the $z$-axis for $T_2$ dephasing)~\cite{BasilewitschPRR20}.

A related set of control protocols that seek to avoid decoherence is summarized under the name of ``dynamical decoupling''~\cite{SuterRMP16}. These protocols use static or time-dependent fields that are designed to average the effect of the interactions with the environment to zero. One way to think of it is as a dynamic way to impose decoherence-free subspaces. Dynamical decoupling can be combined with quantum optimal control to address potentially limiting assumptions on e.g. pulse shapes or pulse timings.  

\subsection{Control of open quantum systems: Exploiting the environment for control}
\label{subsec:open:exploit}

\begin{figure}[tbp]
  \centering
  \includegraphics[width=0.75\linewidth]{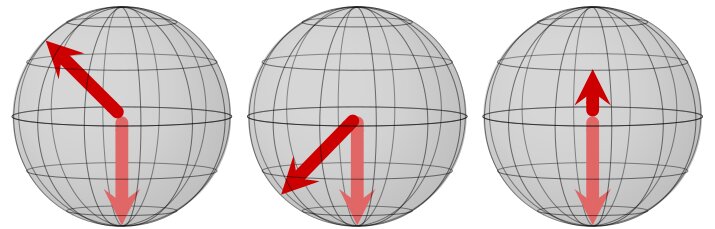}
  \caption{\textbf{Qubit reset} is a paradigmatic example of control problems which can only be solved in the presence of an environment since they require export of entropy. For qubit reset, the protocol should yield the desired target state (light-red arrow) irrespective of the current qubit state (dark red) which may require state rotations (left, middle) and purity increase (right).}
  \label{fig:qubit-reset}
\end{figure}

While beating decoherence may be the most obvious control strategy for open quantum systems, it reaches its limits when the control task requires the presence of an environment. A prominent example is cooling or, more generally, all processes in which the system entropy shall be decreased, see Fig.~\ref{fig:qubit-reset} for illustration. The environment is then needed as an entropy sink. Typical questions for control are how to enhance the cooling rate or how to improve the fidelity of pure state preparation.

Quantum optimal control can be used to answer these questions in non-trivial settings, for example, when the environment is structured, consisting of both strongly and weakly coupled degrees of freedom~\cite{BasilewitschNJP17}.
The strongly coupled modes allow for fast reset, whereas the remainder of the environment can be used for thermalization on a slower timescale. The simplest model to formalize this picture is a qubit strongly coupled to a single auxiliary two-level system (TLS) that undergoes slow decay. The time evolution of the joint system of qubit and auxiliary TLS is then described by a GKLS master equation~\eqref{eq:GKLS}, and compatability with the second law of thermodynamics is ensured by imposing a clear separation between the couplings and the energy scale of the system qubit~\cite{BasilewitschNJP17}. 
%, and guide also the design of analytical solutions. 
We have first used numerical optimization and found pulses that drive the qubit into resonance with the auxiliary TLS in such a way that qubit and TLS have their states exchanged  at the final time. The reset error for the qubit is determined by the temperature of the TLS which is initially assumed to be in a thermal state, and the minimal time for the reset is given by the qubit-TLS coupling strength~\cite{BasilewitschNJP17}. This reset strategy has been implemented experimentally in a slightly modified way for a superconducting qubit, replacing the TLS by a cavity mode~\cite{MagnardPRL18}. 

When adding correlations between qubit and TLS to the initial state, we have found, to our surprise, the final error to be smaller and the resonant population swap to be faster than without initial correlations~\cite{BasilewitschNJP17}. To understand this result, we have derived an approximate analytical model, which captures only the dynamics of qubit and TLS and neglects the TLS's slow decay. Invoking the rotating-wave approximation, which is relevant for details of the dynamics but not the overall control strategy, we have represented the joint state of qubit and TLS in terms of 16 real variables. Their equations of motion  represent a vector flow and can be decoupled by a suitable variable transformation. As a result, the dynamics of the three variables determining the qubit purity, i.e., qubit ground state population and real and imaginary parts of the qubit coherence, can be visualized geometrically~\cite{BasilewitschNJP17}. The visualization reveals that the control rotates initial correlations into qubit ground state population, thereby increasing the final purity and reducing the error. The geometric picture also explains the faster reset: The minimal duration for the simple population swap is $\pi/(2J)$, where $\pi$ is the angle of rotation that the qubit ground state population needs to undergo and $J$ is the strength of the coupling between qubit and auxiliary. Initial correlations reduce this angle. Unfortunately, in order to take advantage of faster and more accurate qubit reset, one would need to know the initial correlations between the system and environmental degrees of freedom which is typically not the case. In contrast, the simple reset via population swap between qubit and auxiliary works irrespective of the initial state~\cite{BasilewitschNJP17}. 

In our initial model for qubit reset, we have assumed an $\sigma_z$-drive on the qubit and a $\sigma_x\otimes\sigma_x$-coupling between qubit and auxiliary~\cite{BasilewitschNJP17}. One may wonder whether a $\sigma_x$- or $\sigma_y$-drive on the qubit would yield even faster reset times. Moreover,  while a  $\sigma_x\otimes\sigma_x$-coupling seems quite natural for population exchange, other couplings could also be engineered, at least for superconducting qubits. We have thus carried out numerical optimizations for a number of couplings~\cite{BasilewitschPRR21}. Their results supported the conjecture that, given any qubit-auxiliary coupling excluding those with $\sigma_z$ for the auxiliary, a specific drive $\epsilon(t)\sigma_i\otimes \openone$ exists that enables qubit reset to the maximally attainable purity. Moreover, the minimum reset duration can only be achieved for qubit and auxiliary on resonance and $\epsilon(t)$ constant~\cite{BasilewitschPRR21}.

The conjecture can be proven combining the tools from controllability analysis introduced in Sec.~\ref{subsec:controllability:basics} with the so-called Cartan decomposition; ; this decomposition is fundamental to classifying all two-qubit operations in the Weyl chamber, cf. Fig.~\ref{fig:weylchamber}. The proof formalizes the ability of purifying a qubit via the condition that the qubit dynamics must be non-unital. This, in turn, requires the dimension of the non-local Cartan subalgebra to be equal or larger than two~\cite{BasilewitschPRR21}. Provided coupling and drive are chosen to fulfill this condition, there exists an optimal choice for the qubit control  $\sigma_i$, such that the reset is time-optimal with $T_{min}=\pi/(2\eta)$, where $\eta$ depends on the choices for the coupling $\sigma_n\otimes\sigma_m$ and the drive $\sigma_i$. Several such choices exist that result in $\eta=J$, the overall shortest possible reset time~\cite{BasilewitschPRR21}. Interestingly, our result saturates the bound obtained from the so-called Time-Optimal Tori theorem~\cite{KhanejaPRA01}.

It is natural to ask whether our results on the minimum duration and maximum purity attainable in qubit reset are modified when the qubit is coupled to an auxiliary degree of freedom with a Hilbert space dimension  larger than two. While the minimum reset time stays the same, the maximally attainable purity increases with increasing Hilbert space dimension~\cite{BasilewitschPRR21}. Indeed, we were able to show rigorously that the purity error, $1-\tr\{\hat\rho_S^2\}$, can be made smaller than some prespecified error tolerance $\varepsilon$ if at least $\lceil d_{\mathrm{B}} (d_{\mathrm{S}}-1)
/ d_{\mathrm{S}} \rceil$ eigenvalues of the initial auxiliary state $\hat\rho_B$ are below $\varepsilon /(2 d_{\mathrm{B}}(d_{\mathrm{S}}-1))$ where $\lceil \cdot \rceil$ denotes the ceiling function, and  $d_{\mathrm{S}}$ and $d_{\mathrm{B}}$ are the Hilbert space dimensions of system (``qubit'') and bath (auxiliary).
The better purification can be rationalized in terms of a unitary operation that acts on both system and auxiliary, reshuffling the spectrum of the joint separable state~\cite{BasilewitschPRR21}. Unfortunately, this improved bound on the reset fidelity does not easily translate into a better reset protocol, as the  reshuffling unitary is likely dependent on the initial state of both system and auxiliary. This leads to the open question whether an approximate unitary can be constructed that surpasses the purity obtained by swapping the population of qubit and auxiliary two-level system, irrespective of the system state.

Qubit reset is the simplest example of quantum reservoir engineering~\cite{PoyatosPRL96} where application of external fields together with the natural interaction of the quantum system with its environment (the ``reservoir'' or ``bath'') 
steers the system toward its ground state. It illustrates the general principle of quantum reservoir engineering to utilize dissipation to drive a system toward a desired steady state. This state can be the ground state but also a  non-trivial state such as an entangled state~\cite{PoyatosPRL96}. However, for non-trivial target states, the required combination of external fields and dissipative dynamics is often rather complex, making it challenging to solve the problem manually. This is another avenue for quantum optimal control.

\begin{figure}[tbp]
  \centering
  \includegraphics[width=\linewidth]{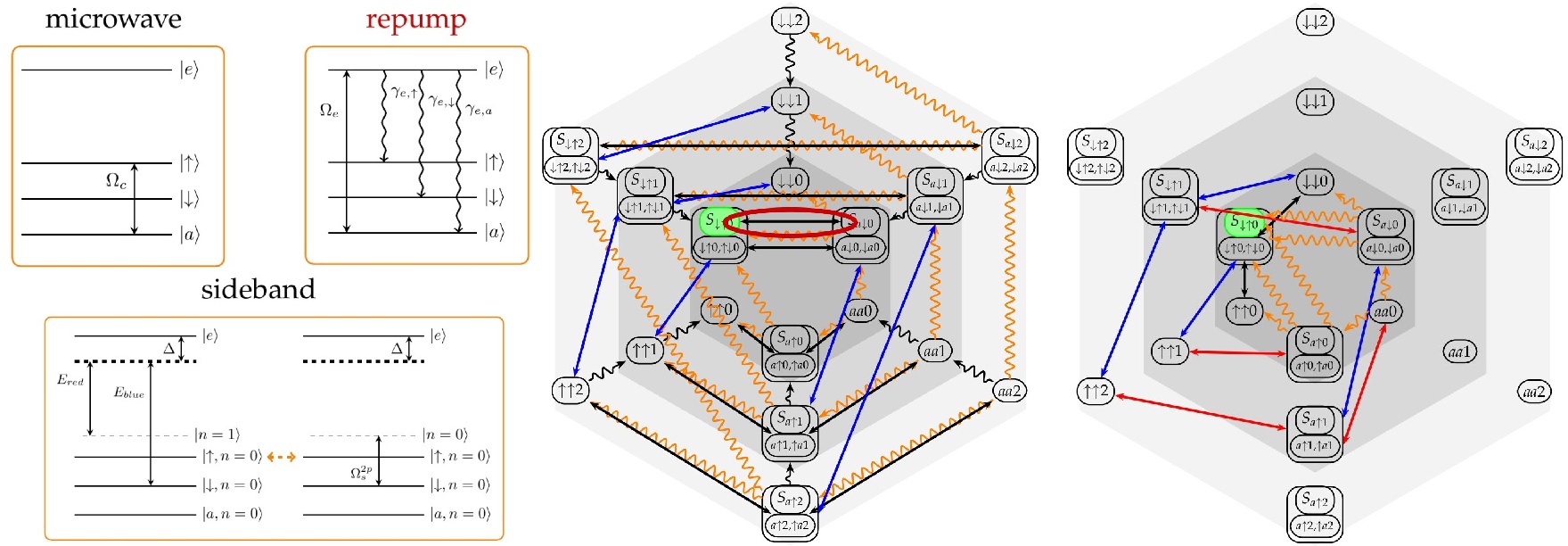}%
  \caption{\textbf{Control of quantum reservoir engineering:} Making the steady state match the desired target state with quantum optimal control~\cite{HornNJP18,ColePRL22}.
    The original protocol~\cite{LinNature13} uses the drives shown on the left, plus laser cooling on two magnesium ions. The population flow induced by these drives is shown in the middle panel with the hexagons formed by the  $\ket{\uparrow}$, $\ket{\downarrow}$, and $\ket{a}$ states of the two beryllium ions, and the size of the hexagons encoding the vibrational excitation in the ion trap. Orange (black) wiggly lines correspond to transitions that are uni-directional due excitation by the repump and subsequent spontaneous emission from the electronically excited state (due to laser cooling), whereas the solid black and blue lines represent transitions that are coherently driven by the microwave and sideband laser. The microwave drives population out of the target singlet state (marked in green). A different microwave transition, identified by optimal control, together with a blue and a red sideband and the repump laser results in uni-direction population flow towards the target state 
(right panel).    Adapted from Ref.~\cite{HornNJP18}.}
  \label{fig:steadystate}
\end{figure}
To illustrate the problem and show how ideas from optimal control can be leveraged to solve it, let us consider the example of preparing two beryllium ions in a maximally entangled, or Bell state~\cite{LinNature13}. The two beryllium ions are co-trapped with two magnesium ions which are laser cooled, ensuring the cooling of all joint vibrations of the four-ion chain. Two hyperfine levels in the electronic ground state of the beryllium ion form the qubit, $\ket{\uparrow}$, $\ket{\downarrow}$, and an additional hyperfine level $\ket{a}$ is also involved in the dynamics, cf. Fig.~\ref{fig:steadystate}. In order to prepare the two qubits in the spin singlet state, a combination of four external drives and spontaneous emission from a very short-lived electronically excited state have been used~\cite{LinNature13}. When all four external fields are kept on continuously, the target state is prepared with a fidelity of about 75\%~\cite{LinNature13}. Using the NLopt package~\cite{NLopt}, we have optimized the parameters of these four fields, i.e., polarizations, Rabi frequencies and detunings, improving the state fidelity to about 90\%~\cite{HornNJP18}. The fidelity is limited because one of the external fields, the microwave drive, couples the target state to another state, as highlighted in the middle panel of Fig.~\ref{fig:steadystate}. In other words, the steady state of the dynamics is not identical to the desired target state. The original protocol balanced the microwave amplitude with respect to the other drives to minimize this error~\cite{LinNature13}. But even with all drive parameters  optimized, the resulting error remains non-negligible~\cite{HornNJP18}. 
We have therefore used optimization to select the best combination of external drives, before optimizing their parameters~\cite{HornNJP18}. To keep the two-stage optimization manageable, we have classified all transitions that can be driven by an external field, reducing the number of possible combinations. As a result of the optimization, a different microwave drive that does not affect the target state is combined with three lasers, cf. right panel of Fig.~\ref{fig:steadystate}, and the polarizations, Rabi frequencies and detunings of all fields are optimized. Then the only remaining source of error is anomalous heating in the trap which determines the final state preparation error, predicted to be  about 90\%, 97\%, resp. 98\% for heating rates of 100$\,$Hz, 10$\,$Hz and 1$\,$Hz~\cite{HornNJP18}.

Optimizing the choice of drives also revealed that laser cooling of the magnesium ions is not essential; instead, a red-detuned sideband transition reduces vibrational excitations in the trap, cf. middle and right panels of Fig.~\ref{fig:steadystate}. This implies the possibility to work with just the two beryllium ions instead of a four-ion chain --- a significant simplification of the  experimental setup! The corresponding reduction in the resources required to prepare a Bell state has meanwhile been demonstrated experimentally by the NIST trapped ion group~\cite{ColePRL22}, with the best state preparation fidelities at 95\%.
The remaining error sources in the modified trapped ion experiment are well understood~\cite{ColePRL22}, suggesting that our theoretically predicted fidelites are indeed feasible.

\begin{figure}[tbp]
  \centering
  \includegraphics[width=0.8\linewidth]{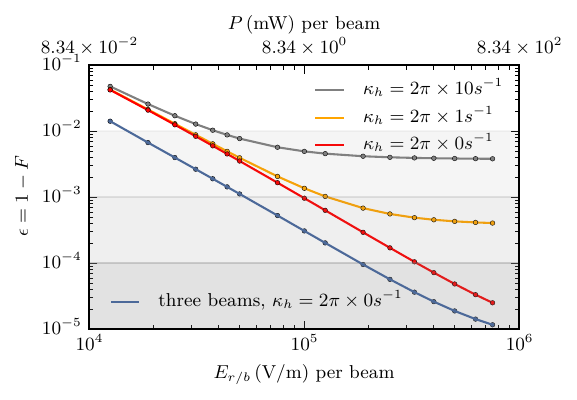}
  \caption{\textbf{Performance bounds from quantum optimal control:} Ultimately attainable error predicted for disspative state preparation 
    as a function of the maximal amplitude of the sideband lasers $E_{r/b}$ (bottom x-axis), resp. laser power (top x-axis), for different heating rates. Adapted from Ref.~\cite{HornNJP18}.
        }
	\label{fig:trappedions-error}
\end{figure}
Both the gain in performance and the reduction of the resources nicely illustrate the utility of optimal control. We can go one step further and ask: What is the ultimately attainable error? In the estimates quoted above, we had capped the laser amplitudes at values that were typical for the experiment of Ref.~\cite{LinNature13}. Assuming that advances in technology eventually will allow for more powerful lasers, even in the UV regime, we can lift this limitation. Large pulse amplitudes allow for operating with larger detunings which is advantageous to minimize undesired photon scattering. Allowing for larger amplitudes, can the state preparation error be pushed below $10^{-4}$? This is the threshold below which error correction is expected to become useful. The question is answered in Fig.~\ref{fig:trappedions-error} which shows the ultimately attainable error as a function of available laser power, accounting also for anomalous heating: For heating rates of 1$\,$Hz and larger, it is not possible to reach the  $10^{-4}$ limit, irrespective of the available power, as seen in the grey and yellow curves levelling off. If anomalous heating can effectively be suppressed on the timescale of the state preparation protocol, the error can be pushed  below $10^{-4}$, cf. red and blue curves in Fig.~\ref{fig:trappedions-error}. The red curve assumes four beams are required to realize the two sideband transitions, whereas the blue curve utilizes a more efficient configuration, achieving the two sideband transitions by combining three beams~\cite{HornNJP18}. Figure~\ref{fig:steadystate} illustrates how quantum optimal control determines an ultimate performance bound, specifically the minimally achievable error. Crucially, it simultaneously identifies the required experimental conditions, here constraints on laser power and heating rates.

\subsection{A more systematic approach to finding suitable interactions: Controllability analysis}
\label{subsec:controllability}
In the previous example, numerical optimization was employed to determine the drives which, in conjunction with spontaneous emission from a short-lived excited state, drive the two-ion system into the desired target state. In formal terms, this corresponds to a search over the space of operators. Such a search incurs a significant numerical cost; even for a system as small as two three-state ions and one vibrational mode, it necessitated a pre-classification of the transition operators. This approach is clearly limited to a few degrees of freedom and cannot be scaled up,  raising the question whether alternatives to brute force optimization over operator space exist.

Ensuring that the steady state of a driven-dissipative evolution matches a desired target state is reminiscent of the controllability analysis introduced in Sec.~\ref{subsec:controllability:basics}. Controllability determines whether a certain target state or target operation can be realized, at least in principle (that is, without bounds on resources such as time, pulse power, or pulse bandwidth).
Traditionally, this analysis has been applied to a fixed Hamiltonian with a given  choice of drives, but it is natural to extend the inquiry to determine which drives are needed to make a system controllable~\cite{LeibscherCommPhys22}. Unlike numerical optimization, which performs a global search over the operator space, controllability analysis is based on the dynamical Lie algebra, exploiting information from the commutators, and may thus provide a more systematic approach to finding the necessary drives. 

Unfortunately, like brute-force numerical optimization, controllability analysis is hampered by the curse of dimensionality. This limitation is immediately apparent as determining evolution-operator controllability requires checking the rank of the dynamical Lie algebra, cf. Sec.~\ref{subsec:controllability:basics}. The most straightforward way to calculate the rank is by explicitly constructing the dynamical Lie algebra from the commutators. However, this is hampered by both the general double-exponential scaling of the operator space and the numerical difficulty to orthogonalize large matrices. Two recent advances address these problems. The first harnesses graph theoretical methods to avoid constructing an orthogonal basis of the dynamical Lie algebra; ultimately, it remains hampered by the double-exponential scaling of the operator space~\cite{GagoQST23}. For general quantum-mechanical time evolutions, the only path to resolving the scalability issue is to perform at least part of the calculations on a quantum device. This strategy has been adopted for controllability analysis by leveraging a hybrid quantum-classical algorithm~\cite{GagoDEA23}.  

\begin{figure}[tbp]
  \centering
  \includegraphics[height=0.25\linewidth]{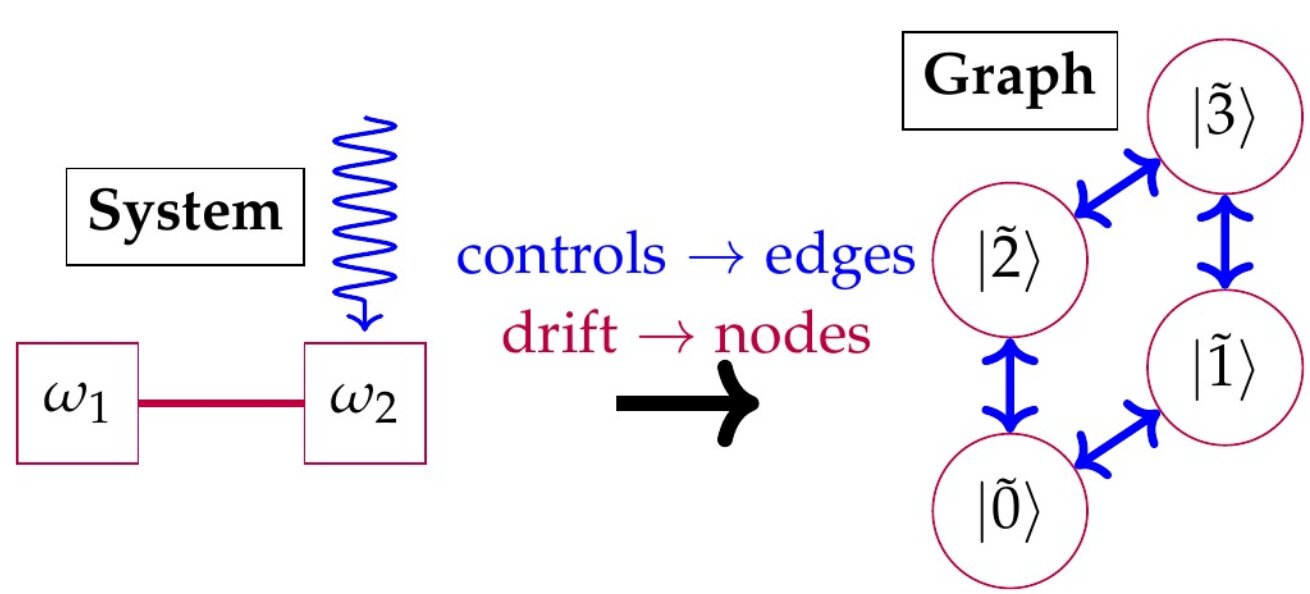}%
 \hspace*{10ex}\includegraphics[height=0.25\linewidth]{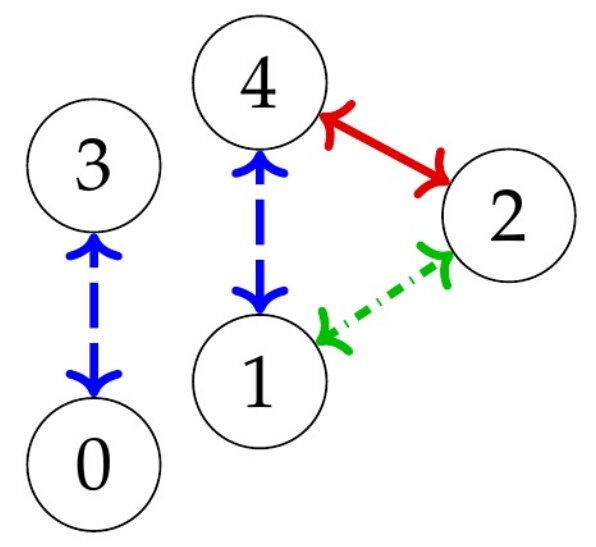}%
 \caption{\textbf{Controllability analysis via graph representation.}
   Left: Mapping a system to a graph for the example of two coupled qubits subject to one external control. The control connects all 
    eigenstates of the total drift (including the qubit-qubit coupling), represented as the nodes of the graph. However, controllability requires, on top of connectedness, all transitions to be decoupled, such that they can be addressed separately. Right: Once the controllability problem is represented as a graph, commutators can be calculated graphically, avoiding explicit construction of the Lie algebra. Here, the red transition is obtained as commutator of blue and green transitions.}
  \label{fig:graph-controllability}
\end{figure}
Representing the controllability problem as a graph makes its solution very intuitive. Separating the Hamiltonian into drift $\hat H_0$ and interactions with external controls, $\hat H(t)=\hat H_0 + \sum_ju_j(t)\hat H_j(t)$, the eigenstates of the drift are taken to be the nodes of the graph, whereas the controls become edges connecting the nodes. As an example, the left part of Fig.~\ref{fig:graph-controllability} shows the graph for two coupled qubits, one of which is driven by an external control. A system is controllable if a connected subgraph exists that contains all nodes of the graph and only decoupled\footnote{Two transitions are coupled if they are connected to the same control $u_j(t)\hat H_j$ and have the same transition frequency.} transitions~\cite{GagoQST23} (and refs. therein). The two examples shown in Fig.~\ref{fig:graph-controllability} are \textit{not} controllable: In the example on the left, the subgraph contains all nodes but the transitions are coupled (indicated by the color). In contrast, the right-most subgraph contains only decoupled transitions but not all nodes of that example are contained in the subgraph. The examples in  Fig.~\ref{fig:graph-controllability} have mainly an illustrative purpose. For the important case of coupled qubit arrays,  Ref.~\cite{GagoQST23} provides algorithm flowcharts for setting up the graph,  constructing all commutators graphically and checking the subgraph controllability condition. The utility of the graphical controllability test was shown for qubit arrays consisting of five qubits connected in T-shape as in the ibmq\_quito architecture~\cite{GagoQST23}.  In the original design, each qubit is driven by a $\hat\sigma_x$-control, and neighboring qubits are coupled via $\hat\sigma_x\hat\sigma_x+\hat\sigma_y\hat\sigma_y$. Controllability analysis revealed that the number of external controls can be reduced from five to two, keeping the qubit-qubit coupling fixed. Using instead qubit-qubit couplings of the type $\hat\sigma_x\hat\sigma_x+\hat\sigma_y\hat\sigma_y+\hat\sigma_z\hat\sigma_z$, a single external control is sufficient for evolution-operator controllability, i.e. for implementing arbitary unitaries on the five qubits~\cite{GagoQST23}. Since external controls come with considerable overhead in quantum chip design, reducing their number may be beneficial in view of scaling up the size of those chips. For larger and larger devices, however, running controllability tests on classical computers becomes impractical due to the exponential scaling of the Hilbert space. 

To enable controllability tests for large qubit arrays, a hybrid quantum-classical algorithm based on a parametric quantum circuit was devised~\cite{GagoDEA23}. The algorithm hinges on the proof that controllability is linked to the number of independent parameters $\theta_i$ of the circuit~\cite{GagoDEA23}. These parameters are, in turn, obtained by dimensional expressivity analysis, which was originally developed for obtaining a maximally expressive circuit ansatz with a minimum number of parameters~\cite{FunckeQuantum21}. Crucially, dimensional expressivity analysis can efficiently be implemented on quantum hardware~\cite{FunckeQuantum21}.

\begin{figure}[tbp]
  \centering
  \includegraphics[width=0.7\linewidth]{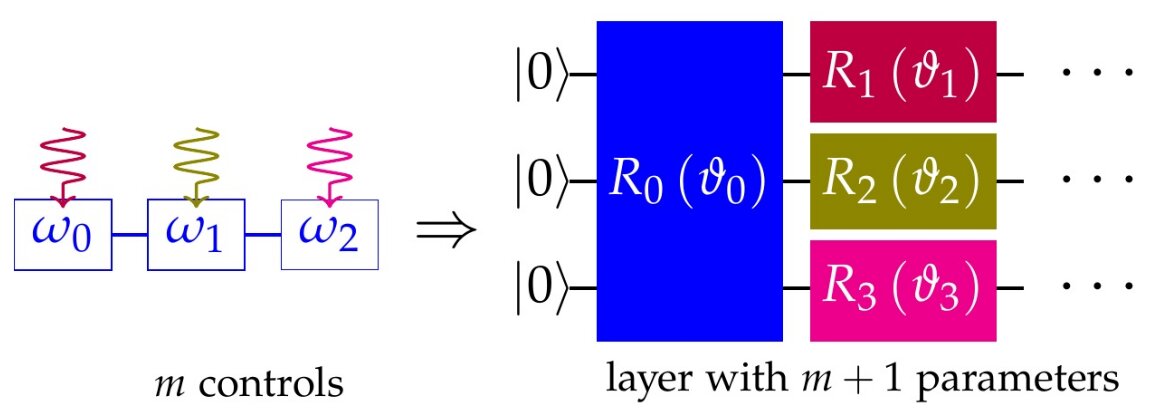}
  \caption{
\textbf{Controllability analysis via a hybrid quantum-classical algorithm:}
    Parametrized quantum circuit to determine pure state controllability for three qubits. Adapted from Ref.~\cite{GagoDEA23}.}
  \label{fig:DEA-controllability}
\end{figure}

The hybrid quantum-classical algorithm to determine controllability proceeds as follows. The controllability problem is presented as a parametric quantum circuit by associating 
each control plus the drift with rotation angles $\theta_j$. One layer of the circuit is shown in Fig.~\ref{fig:DEA-controllability}, where $R_j(\alpha)=\exp(i\alpha \hat H_j)$ with $j=0,\ldots, m$ for $m$ controls. The complete circuit consists of $n_l$ layers with different values of the parameters $\theta_j$ in each layer. Dimensional expressivity analysis for the circuit~\cite{FunckeQuantum21} is carried out. 
If the circuit reaches maximal expressivity, the system is controllable (in the sense that any pure state of the qubit array can be reached), otherwise another layer is added and the dimensional expressivity test is repeated~\cite{GagoDEA23}. In case the controllability test fails even with a large number of layers, different initial states and random choices of the parameters $\theta_i$ need to be tested to confirm the negative controllability result.

Typically, evolution operator controllability is the focus (rather than pure state controllability), as it is a requirement for universal quantum computation. Similarly to extending quantum state tomography to quantum process tomography, this extension of the  controllability test is achieved  via the Choi-Jamio\l{}kowski isomorphism~\cite{GagoDEA23}. It implies using twice as many qubits which must be initially prepared in an entangled state. Aside from this overhead, the algorithm proceeds as before.

Implementing the algorithm on an actual device will enable controllability tests to be used for resource-efficient design of quantum chips. The controllability analysis provides the minimal number of local controls and qubit couplings required for controllability and, consequently, universal quantum computation~\cite{GagoDEA23}.  
It allows for obtaining this information before a device is actually built, provided the associated quantum circuit can be implemented on a different device. However, reducing the number of controls comes at the expense of longer gate durations. Therefore, balancing  the number of external controls and the gate durations will be essential. To achieve this balance, it will be important to understand how the removal of redundant controls affects the minimum gate duration.

\subsection{Control of open quantum systems: Creating the desired dissipation}
\label{subsec:open:create}

\begin{figure}[tbp]
  \centering
  \includegraphics[width=0.6\linewidth]{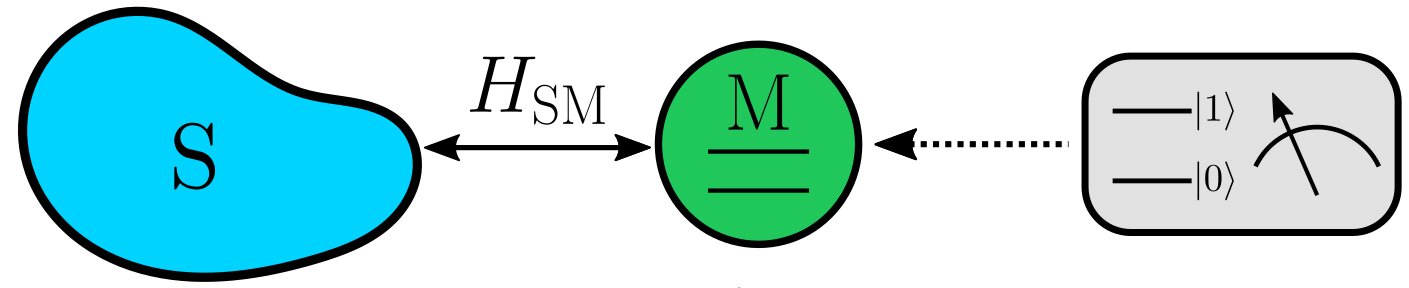}
  \caption{\textbf{The principle of measurement-induced dynamics:}
Coupling a system to one or more auxiliary degrees of freedom ($M$ for ``meters'') which are discarded after their interaction with the system $S$, for example by projective measurements. Adapted with permission from Ref.~\cite{LangbehnPhD}.
  }
  \label{fig:system-meter}
\end{figure}

Let us now return to the control of open quantum systems. Below, we will examine examples demonstrating how the tools of controllability analysis identify the necessary operators, as conjectured in Sec.~\ref{subsec:open:exploit} in the context of quantum reservoir engineering. Before proceeding with these examples, we first recap the known control strategies  for open quantum systems. 
Rephrasing the preliminary set of rules discussed on p.~\pageref{list:rules}, we can classify them as follows: 
\begin{enumerate}\setlength{\itemsep}{1ex}
\item Detrimental: Avoiding decoherence\\
  As discussed above in Sec.~\ref{subsec:open:fight},  the impact of the environment is detrimental for Markovian dynamics\footnote{For systems with non-Markovian dynamics, the question of available control strategies is still open. It is more complex than for systems with Markovian dynamics because of additional ways to exploit the environment~\cite{ReichSciRep15}.} and for targets that do not require any entropy export. The available strategies are then to operate fast compared to the decoherence timescale or to engineer protection from decoherence via decoherence-free subspaces or  dynamical decoupling. Quantum optimal control theory is useful to implement these strategies, since it allows for designing control protocols with durations at the quantum speed limit or leveraging decoherence-free subspaces, as we have seen in the examples in Sec.~\ref{subsec:oct} and Sec.~\ref{subsec:open:fight}.
\item Beneficial-1: Exploiting dissipation\\
  For control targets that require entropy export, the presence of the environment is crucial and, consequently, beneficial. The key paradigm is quantum reservoir engineering~\cite{PoyatosPRL96}, which involves choosing external drives such that the steady state of the driven-dissipative evolution matches the desired target state. Section~\ref{subsec:open:exploit} presented the Bell state preparation for trapped ions as an example, demonstrating how optimal control theory can simplify the experimental design, improve state preparation errors, and identify ultimate performance bounds. 
\item Beneficial-2: Creating desired dephasing and dissipation\\
  When the required dissipation channels are not naturally available, they can be effectively created by coupling the system to auxiliary degrees of freedom. Following their interaction with the system, these auxiliaries are either simply discarded or subjected to projective measurements, see Fig.~\ref{fig:system-meter}. This dissipation engineering can be viewed as quantum reservoir engineering 2.0. It offers the advantage of being readily applicable to many-body systems and has, consequently,
  become a crucial component of the quantum control toolbox, with applications in quantum simulation and quantum error correction~\cite{HarringtonNRP22}.
\end{enumerate}
The remainder of this section will present three examples of engineered decoherence and dissipation, illustrating the basic concepts and showcasing where tools from quantum control can be leveraged.

The first example exploits dephasing to endorse adiabaticity~\cite{MenuPRR22,SveistrysQuantum25}. The basic intuition can be rationalized with the Landau-Zener model for an avoided crossing, see also Fig.~\ref{fig:adia} (bottom left),
\[
\hat H_S(t)=\frac{\epsilon t}{2}\hat\sigma_z+\frac{g}{2}\hat\sigma_x\,.
\]
The goal is to avoid diabatic transitions. One strategy is to adiabatically traverse the crossing, but often such slow dynamics cannot be realized. Diabatic transitions would also be avoided if the coherence between the two levels is suppressed. This corresponds to pure dephasing of the Landau-Zener qubit. 
One possibility to engineer this desired dephasing is by coupling the qubit to an auxiliary degree of freedom, e.g. a cavity coupled to a thermal bath, which we term the ``meter'' to emphasize the analogy with quantum measurements~\cite{MenuPRR22}. Pure dephasing is achieved via a system-meter coupling of the same form as quantum non-demolition (QND) detection. This requires the system-meter coupling to commute with the system Hamiltonian. Since $\hat H_S(t)$ is time-dependent, the system-meter coupling must also be time-dependent. Assuming this coupling can be engineered and made sufficiently strong, diabatic transitions in the Landau-Zener qubit can be strongly suppressed~\cite{MenuPRR22}. A more realistic scenario, however, involves a stroboscopic, rather than fully time-dependent implementation of the system-meter coupling; in this case the effectiveness of suppressing diabatic transitions is governed by the coupling rate~\cite{MenuPRR22}.

Applying engineered dephasing to many qubits, such as those found in quantum annealers, requires the system-meter coupling to be both time-dependent and global~\cite{SveistrysQuantum25}. Such coupling provides speedups in the time-to-solution of the annealing that are linearly proportional to the coupling strength. 
Surprisingly, replacing the cavity "meter" with a two-level system --- which allows for calculating the exact reduced system dynamics --- reveals that a fully coherent mechanism (effective energy rescaling) is the more effective approach for enhancing the system's adiabaticity, rather than pure dephasing~\cite{SveistrysQuantum25}.
Implementing the required system-meter couplings in quantum annealing is daunting and  beyond what is readily available in existing devices. This difficulty can be mitigated
because the actual schedule is often irrelevant, provided it yields the final Hamiltonian that encodes the computational problem. This fact can be exploited to simplify the time-dependence of the system-meter coupling, albeit at the cost of smaller speedups~\cite{SveistrysQuantum25}. In summary, while conceptually very appealing, the implementation of engineered dephasing protocols is rather difficult, at least in time-dependent scenarios such as quantum annealing.
Therefore, applications that require only static system-meter interactions may represent  a more promising way forward.

Static system-meter interactions can also be used to engineer dissipation, or energy relaxation, rather than pure dephasing. The driven-dissipative dynamics then effectively ``cools'' the system into the target state.
An important application of this approach is to overcome the limitations of quantum reservoir engineering protocols that are based on natural decay processes. Overcoming these limitations would be crucial for realizing theoretical proposals targeting the preparation of non-trivial many-body quantum states as steady states of Markovian master equations~\cite{KrausPRA08}.
To understand the limitations of quantum reservoir engineering for many-body systems, consider the example of trapped ions, discussed Sec.~\ref{subsec:open:exploit}. There, the dissipative process is spontaneous emission from an electronically excited state, which presents several drawbacks. (i) Spontaneous emission is essentially local and cannot, by itself,  not introduce entanglement. (ii) It leads to leakage to undesired states. Even for just two ions, this necessitated a complex combination of several fields to repump all population into the cooling cycle. (iii) Finally, a natural decay process such as spontaneous emission, which can be coupled to in a controlled way, may simply not exist for a given target state or space.

Instead of relying on natural decay processes, measurement-induced dynamics~\cite{HarringtonNRP22} exploits the fact that quantum measurements induce non-unitary evolution in the measured system~\cite{WisemanBook}. Continuous measurements, in particular, can be described by a GKLS master equation, cf. Eq.~\eqref{eq:GKLS}, where the system part of the system-meter interaction Hamiltonian defines the jump operators~\cite{WisemanBook}.
In other words, quantum measurements provide a constructive recipe to engineer jump operators at will~\cite{HarringtonNRP22}! A complementary perspective on measurement-induced dynamics is offered by treating the meter degrees of freedom as ``colliders'' in so-called collision models~\cite{CiccarelloPhysRep22}, which highlights the effective ``simulation'' of the desired open system dynamics~\cite{CattaneoPRL21}.

To illustrate the approach with a specific example, consider the so-called AKLT model\footnote{The model is named after Affleck, Lieb, Kennedy and Tasaki who introduced it~\cite{AKLT1,AKLT2}.} describing a chain of spin-1 particles with only nearest-neighbor interactions. The model is frustration-free; with periodic boundary conditions, it has  a non-degenerate ground state that is a paradigmatic example of both a matrix product state and a symmetry-protected topological phase~\cite{AKLT1,AKLT2}. This ground state can be prepared dissipatively, without any action of the Hamiltonian~\cite{RoyPRR20}. To this end, each link in the chain of spin-1 particles is repeatedly connected to five meter qubits. Following their interaction with the chain, they are discarded and replaced by freshly initialized qubits. The five jump operators resulting from the system-meter interactions deplete any population in excited statess~\cite{RoyPRR20}. A second possibility to prepare the AKLT ground state is by alternating unitary evolution and fusion measurements~\cite{SmithPRXQ23}. In both cases, the number of meters scales with the size of the AKLT chain. Remarkably, control theory allows for modifying the protocol of Ref.~\cite{RoyPRR20}, reducing the number of links that are cooled to a single one~\cite{LangbehnPRXQ24}. This requires the action of the AKLT Hamiltonian, in addition to coupling one  link to five meter qubits. It comes at the expense of slower convergence to the target state with increasing size of the chain. Meter qubits are a precious resource, making  the modification very appealing in terms of resource efficiency --- the repeated preparation of a large number of meter qubits implies a significant operational cost, whereas engineering the two-body couplings of the AKLT Hamiltonian is of a similar difficulty as engineering the coupling to the meters.

\begin{figure}[tbp]
  \centering
  \includegraphics[width=\linewidth]{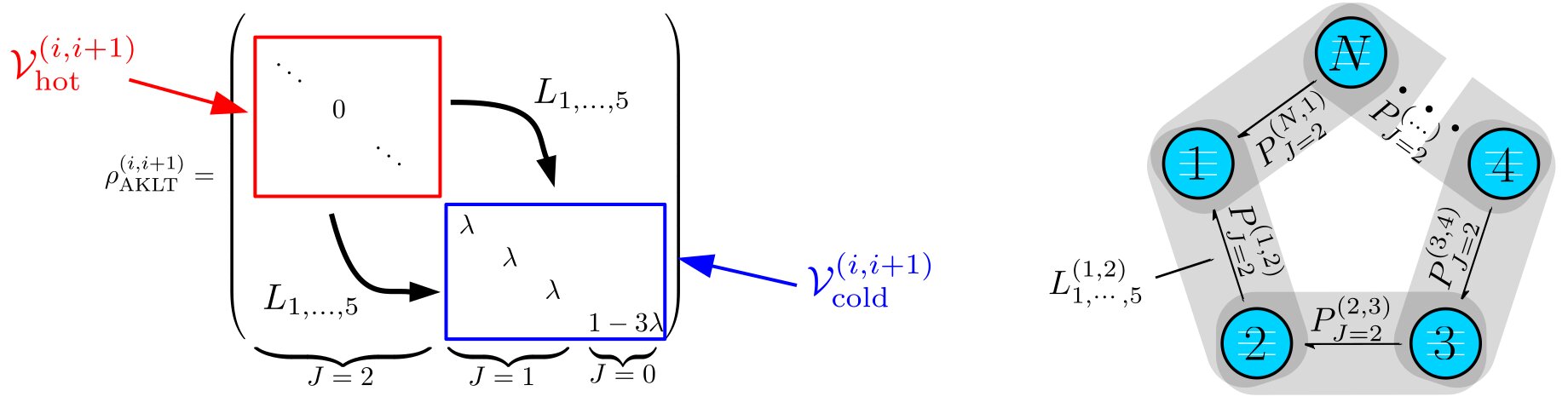}
  \caption{\textbf{Dilute cooling of the AKLT chain:} Weak measurements of a single link of $N$ spin-1 particles, coupled by nearest-neighbor interactions, allow one to drive the system into its ground state. The couplings to the five meter qubits give rise to jump operators $L^{(1,2)}_{1,\ldots,5}$ which 
    ensure that all population in the local hot subspace is cooled, whereas the interactions within the chain move all excitations through the chain until they reach the cooled link. 
    Adapted from Ref.~\cite{LangbehnPRXQ24}.}
  \label{fig:dilute-cooling}
\end{figure}
The possibility of such "dilute" cooling can be rationalized by combining two concepts,  analysis of the state space structure and the notion of controllability (cf. Section~\ref{subsec:controllability:basics})~\cite{LangbehnPRXQ24}.
To this end, consider the reduced state on the cooled link (sites 1,2 in Fig.~\ref{fig:dilute-cooling}); it is obtained by integrating out all other sites ($3,\ldots, N$ in Fig.~\ref{fig:dilute-cooling}).
%It is defined on the local Hilbert space of the cooled link. 
Denoting the target state defined on all of the chain by $\rho_\oplus$, we refer to the support of $\rho_\oplus$ on the cooled link  as the cold subspace, and its complement in the local Hilbert space (of sites 1,2) as the hot subspace, cf. Fig.~\ref{fig:dilute-cooling}. The interaction with the meter qubits is designed in such a way that all population in the hot subspace is transferred into the cold subspace, without affecting any population in the cold subspace. Dilute cooling works because all excited states of the AKLT chain have a non-zero overlap with the local hot subspace on the cooled link: Cooling depletes the population of the local hot subspace, but the interactions within the chain refill it with excitations from other parts of the chain, until all excitations have been cooled. Therefore, the minimal overlap of the excited states with the hot subspace  defines the cooling rate~\cite{LangbehnPRXQ24}. 

The non-vanishing overlap of all excited states with the hot subspace is a beneficial characteristic of the AKLT model but this intuition can be formalized into  necessary and sufficient conditions~\cite{LangbehnPRXQ24}. This allows dilute cooling to be generalized beyond the AKLT model. A first necessary condition is that the target state, when reduced to the cooled link, must be distinguishable from the reduced excited states; in essence, the local hot subspace cannot be empty, $V_{\text{hot}}^{\left(i,i+1\right)}  \neq\emptyset$. 
A second necessary conditions concerns the coherent interactions within the system. To formulate this condition, we construct a vector space containing all quasi-local operators that respect the target state as elements. This space has been dubbed the ``kernelizer'' since  the target state is an eigenstate of these operators with eigenvalue zero~\cite{LangbehnPRXQ24}. 
The condition then dictates that the target state must necessarily be the only state invariant to both the cooling mechanism (i.e., the jump operators) and all coherent interactions originating from the kernelizer. This second necessary condition ensures that excitations on links that are not directly cooled propagate through the system until they reach the cooled link.
Since the elements of the kernelizer are not necessarily part of the system Hamiltonian, the condition is only necessary, but not yet sufficient.
It can be turned into a sufficient condition by ensuring controllability on the complement of the target subspace. In other words, if the Lie algebra generated by the kernelizer  has full rank, coherent interactions exist that, when added to the system Hamiltonian, ensure the necessary population flow towards the cooled link~\cite{LangbehnPRXQ24}.

The necessary condition of local distinguishability of the target state is not fulfilled for certain highly entangled states, such as GHZ states. Indeed, these states  can only be cooled with a number of meters that is extensive in system size.
Conversely, the state space structure of the AKLT model is expected for any frustration-free spin chain with Hamiltonian given as
a sum over noncommuting terms. These systems should therefore be amenable to dilute cooling. This was demonstrated for the example of the Majumdar-Ghosh model, a chain of spin-$1/2$ particles with nearest- and next-to-nearest neighbor interactions. 
Remarkably, a suitable combination of jump operators proved sufficient
to selectively prepare one out of two degenerate ground states~\cite{LangbehnPRXQ24}. 

The key limitation of dilute cooling follows from the necessary conditions: As seen in the example of the GHZ states above, not every target state allows for the required partitioning of the state space. Moreover, the protocol requires the implementation of three-body interactions between the two sites of the cooled link and the meters. Three-body interactions are often engineered in an effective way from two-body interactions; this incurs an additional operational overhead. These limitations can be overcome by stochastic cooling which utilizes only two-body interactions between system sites and meters (the parameters of which are chosen randomly). Stochastic cooling does not make any assumptions on the target state except for being sufficiently gapped~\cite{Langbehn25}. The approach is universal in the sense that it requires no detailed knowledge of the Hamiltonian or its spectrum. This generality requires a trade-off in efficiency: it necessitates many meter qubits prepared in their ground state, resulting in a large resource overhead~\cite{Langbehn25}.

The examples for engineered dissipation discussed above are based on meters that are simply discarded after their interaction with the system. Protocols that read out the state of the meter  before disposing them are referred to as active. The corresponding variant of quantum control is quantum feedback~\cite{WisemanBook}. While beyond the scope of this lecture, the combined use of coherent interactions, quantum measurements, and quantum feedback, also referred to as ``quantum interactive matter'', 
is an exciting emerging avenue of quantum control. 

\section{Open questions}
\label{sec:open}

With this lecture I have attempted to forge a link from the very basic concepts of  quantum control theory, as established in the 1980s and 1990s, to present-day applications in quantum information science. The sheer number of the latter is growing at a breathtaking speed~\cite{GlaserEPJD15,KochEPJQT22}, making quantum optimal control an indispensable tool in the practical implementation of quantum technologies. Moreover, the range of applications keeps expanding, as highlighted by the most recent examples from dissipation engineering in many-body systems~\cite{HarringtonNRP22}. It would therefore be presumptuous to predict the future directions of quantum control.
Instead, I will conclude this lecture  with a brief overview of open questions that I personally find interesting, challenging, or otherwise relevant.
The discussion so far as well as my concluding remarks have deliberately adopted a focused perspective illustrating key concepts of quantum control and providing intuition with examples drawn primarily from the work of my research group.

Frequent questions that are raised after my lectures concern the robustness  of control protocols as well as the comparison of machine learning to optimal control. While a detailed discussion is found in Ref.~\cite{KochEPJQT22}, let me summarize the key points. 
Robustness, as any property of a control, needs to be accounted for in the design of the protocol. In quantum optimal control, the most straightforward way is to include a desired feature in the target functional or to optimize over an ensemble of systems representing parameter fluctuations. However, these techniques are quite costly and often do not work ``out of the box''. More practical approaches to include robustness would significantly improve the  applicability of quantum optimal control. Moreover, it would be important to quantify whether robustness can be achieved for operations at the quantum speed limit~\cite{BasilewitschChaotic23} or whether one needs to balance robustness and operation speed.

Quantum optimal control and machine learning are closely linked~\cite{KochEPJQT22}. For example, both need to quantify their targets in terms of suitable figures of merit, with a lot of potential for cross-fertilization. Both are hampered by the cost of solving the quantum mechanical equation of motion. They differ in their strengths and weaknesses. For example, quantum optimal control often has superior convergence properties but its control protocols are not readily applicable to slightly different settings. Machine learning is designed to be transferrable but is hampered by the cost of quantum measurements --- quantum physics is not big data. A  comprehensive and fair comparison of the two approaches is another open challenge.  

An attentive reader will have wondered why controllability has been discussed primarily for unitary evolution whereas control protocols focused on open quantum systems. Controllability of open quantum systems is largely uncharted territory! A key limitation for analyzing the controllability of open quantum systems is that traditional controllability analysis does not allow for resource constraints, including time. But open quantum systems are characterized by several competing timescales. If these need to be ignored, only asymptotic statements are possible. This explains why controllability results for open quantum systems have largely been confined to the characterization of reachable sets of states in Markovian dynamics~\cite{ToSHRMP09}. In order to overcome this limitation, one would need to find a statement of the controllability problem that accounts for competing timescales. A second difficulty is given by the lack of a general equation of motion for open quantum systems. It is thus not clear how to rigorously state the controllability problem for systems with non-Markovian dynamics. These challenges call for new ideas. They will likely emerge from an overall better understanding of open systems control and the fast-paced development of quantum engineering.

Our current understanding of control strategies for open quantum systems clearly is incomplete. An attentive reader will have noticed a missing discussion of control strategies for non-Markovian dynamics in the preliminary balance in Sec.~\ref{subsec:open:create}. Memory effects or strong coupling to environmental degrees of freedom may serve as a resource for control, as showcased in examples such as the implementation of quantum gates thanks to coupling to environmental modes~\cite{ReichSciRep15}. But control design typically requires knowledge of the dynamics, and 
the difficulty to compute non-Markovian dynamics has so far impeded a comprehensive analysis.

Another avenue that is being explored only recently is the interplay of strong driving and dissipation. Take the example of dynamics described by Markovian GKLS master equations. If the system is strongly driven, the master equation must be derived in the dressed basis (with respect to the drive)~\cite{AlbashNJP2012}. The resulting drive-dependence of the jump operators can be leveraged in quantum optimal control, for example to speed up qubit reset~\cite{BasilewitschNJP19}. Exploring the full potential of this type of control will be an important contribution to the field of dissipation engineering and quantum interactive matter. 

Finally, a number of intriguing questions concern the ``classical-quantum boundary'' in the control of quantum systems. Whereas the external fields used in most of this discussion  were classical, the auxiliary degrees of freedom in Sec.~\ref{subsec:open:create} can be viewed as quantum controllers. This perspective becomes particularly relevant when combining coherent control and quantum feedback~\cite{RouchonARC22}. More generally, whether an external field should be modeled as classical or quantum depends on its state. But is a given task better carried out with a classical or a quantum control? Naively, one would assume classical controls to be more readily available. At the same time, operating a quantum device with classical controls incurs both calibration overhead and noise. This problem is exacerbated in quantum feedback control which is based on classical external fields and quantum measurements (with additional costs). In contrast, quantum controllers can operate autonomously and thus minimize the costs associated with the classical-quantum interface. 
While already exploited in autonomous quantum error correction~\cite{MirrahimiNJP14}, its potential for quantum control more generally has not yet been tapped.

\begin{acknowledgement}
  % If you want to include acknowledgments
  I am indebted to my former and present students for dedicating their time, energy, and efforts to advancing quantum control theory. This lecture is largely based on the master's and doctoral theses by Daniel Reich, Michael Goerz, Sabrina Patsch, Daniel Basilewitsch, Karl Horn, Fernando Gago, and Josias Langbehn.
 I would like to thank Simone Sergi for inspiration and encouragement. 
\end{acknowledgement}    

\ethics{Competing Interests}{The author declares no competing interests.}

\printbibliography
%\bibliographystyle{spphys.bst}
%\bibliography{refs}%

\section*{Appendix}

\section{Prerequisites}
\label{app:prerequ}

\subsection{Frame transformations and rotating-wave approximation (RWA): Two-level system}
\label{subsec:frameTLS}
Assuming that we need to consider only the ground and first excited state of an atom, the atomic Hamiltonian, or drift, is given by
\[
\hat H_0=\frac{\omega_0}{2}\hat\sigma_z\,,
\]
and the electric dipole interaction with a laser becomes
\[
\hat H_I(t)= \vec d\cdot\vec E(t) \hat\sigma_x\,,
\]
where $\vec E(t)=E_0 S(t) \cos(\omega_Lt)$. Here, a shape function $S(t)$ has been introduced in order to model the switching on and off of the electric field. Since the electric field in $\hat H_I(t)$ appears as it is realized in the lab, this representation is referred to as "lab frame". 

When we consider resonant excitation of the atom (which is a prerequisite for truncating the large atomic Hilbert space to just two levels), $\omega_0$ and $\omega_L$ differ only by a small amount, the detuning $\Delta_L=\omega_0-\Delta_L$. When oscillations with $\omega_{0/L}$ are much faster than all other relevant timescales, it is useful to transform the time-dependent Schr\"odinger equation, 
\[
\frac{\partial}{\partial t}\ket{\psi(t)} = -i\hat H(t) \ket\psi(t)\,,
\]
into a reference frame that rotates with one of the large frequencies, $\omega_{0}$ or $\omega_{L}$. This is achieved by a unitary transformation $\hat U$ in the usual way, that is, we insert $\ket\psi=\hat U\ket{\psi^\prime}$ and identity into the time-dependent Schr\"odinger equation,
\[
\frac{\partial}{\partial t}\left(\hat U(t)\ket{\psi^\prime(t)}\right) = -i\hat U(t)\hat U^+(t)\hat H(t) \left(\hat U(t)\ket{\psi^\prime(t)}\right)\,,
\]
which can be rewritten for $\ket{\psi^\prime(t)}$,
\[
\frac{\partial}{\partial t}\ket{\psi^\prime(t)} = 
-i\left(\hat U^+(t)\hat H(t)\hat U(t) - \hat U^+(t)\frac{\partial \hat U(t)}{\partial t}\right)\ket{\psi(t)}=
-i\hat{ H}^\prime(t) \ket{\psi^\prime(t)}\,.
\]
The frame rotating with $\omega_0$ is reached when taking $\hat U=e^{-i \hat H_0 t}$. In this case $\frac{\partial \hat U}{\partial t}=-i\hat H_0\hat U(t)$ and $\hat{ H}^\prime(t)$ becomes
\[
\hat{ H}^\prime(t)= \begin{pmatrix} 0 & -\frac{1}{2}\Omega_0 S(t) \left(e^{-i(\omega_L-\omega_0)t}+e^{-i(\omega_L+\omega_0)t}\right)\\
-\frac{1}{2}\Omega_0 S(t) \left(e^{i(\omega_L-\omega_0)t}+e^{i(\omega_L+\omega_0)t}\right) & 0 \\
\end{pmatrix}\,,
\]
where we have introduced the Rabi frequency $\Omega_0=\vec d\cdot\vec E_0$. 
The rotating wave approximation consists in neglecting the terms that oscillate with twice the large frequency,
\[
\hat H_{RWA}^\prime(t) = \begin{pmatrix} 0 & -\frac{1}{2}\Omega_0 S(t) e^{i\Delta_Lt}\\
-\frac{1}{2}\Omega_0 S(t) e^{-i\Delta_Lt} & 0 \\
\end{pmatrix}\,.
\]
This is justified when the fast oscillations average to zero on the timescale of interest. Indeed, all remaining time-dependencies in $\hat H_{RWA}^\prime(t)$ are slow since $\Delta_L\ll\omega_L,\omega_0$ and $S(t)$ has been introduced as the envelope of the electric field amplitude. 

In this simple example, the transformation into the rotating frame is identical to the tranformation into the interaction picture. For $N$-level systems, the two usually do not coincide any longer, as discussed below. In general, several choices for a rotating frame are possible, while there is only one interaction picture. This can be seen already for a two-level system when considering a more general expression for the electric field, $\vec E(t)=\vec E_0 S(t)\cos(\omega_L t+\varphi(t))$ including a time-dependent phase $\varphi(t)$. Such a phase is generated for example when "chirping" the pulse. In order to understand this expression, it is instructive to sketch the electric field as a function of time for $\varphi(t)=0$ and $\varphi(t)=\alpha t^2$ where $\alpha$ is the so-called chirp rate.
The general form of the unitary transformation into a rotating frame for a two-level system is given by
$\hat U(t) = \exp(i/2 \vartheta(t)\hat\sigma_z)$. In the transformation above, $\vartheta(t)$ was taken to be $\vartheta=\omega_0 t$ but equally valid choices are $\vartheta(t)=\omega_L t$ or 
$\vartheta(t)=\omega_L t+\varphi(t)$. With these choices, the frame rotates with the central, respectively the instantaneous, laser frequency.
It is instructive to derive the corresponding Hamiltonians $\hat H_{RWA}^{\prime\prime}(t)$, $\hat H_{RWA}^{\prime\prime\prime}(t)$. It is also instructive to visualize the difference between the frames on the Bloch sphere: The equation of motion for a two-level system can be rewritten in terms of the Bloch vector $\vec r(t)\in \mathbb R^3$ precessing around a vector $\vec\Omega(t)$ representing the Hamiltonian,
\[
  \frac{d}{dt}\vec r(t) = \vec r(t) \times \vec\Omega(t)\quad\mathrm{with}\quad
  \hat H(t) = \vec \Omega(t)\cdot\vec\sigma\,.
\]

\subsection{Frame transformations and rotating-wave approximation for $N$-level systems}
\label{subsec:frameNLS}
A good choice of rotating frame is essential. First of all, it simplifies the equations and provides a more intuitive understanding of the (essential) dynamics. 
Further, and even more importantly, for $N$-level systems with $N>2$ it gives rise to different ways to take the rotating-wave approximation. This is exploited for example when deriving effective Hamiltonians to model superconducting qubits~\cite{GirvinLectureNotes2011}. %% add citation

To see how one can generalize the concept of frame transformation, let us consider a three-level system where levels 1, 2 and levels 2, 3 are connected by a dipole transition (i.e., the corresponding matrix elements in the Hamiltonian are non-zero)~\cite{ShoreBook2011}. The system interacts with a two-color electric field $\vec E(t)=\vec E_1S_1(t)\cos(\omega_1t)+\vec E_2S_2(t)\cos(\omega_2t)$, such that the total Hamiltonian, $\hat H(t)=\hat H_0+\hat H_I(t)$, is given by
\[
\hat H(t) =\begin{pmatrix}
    \mathcal E_1 & \hat{\vec d}_{12}\cdot \vec E(t) & 0 \\
    \hat{\vec d}_{21}\cdot \vec E(t) & \mathcal E_2 & \hat{\vec{d}}_{23}\cdot \vec E(t) \\
    0 & \hat{\vec d}_{32}\cdot \vec E(t) & \mathcal E_3 
\end{pmatrix}
\]
Applying the following unitary transformation to the time-dependent Schr\"odinger equation,\[             \hat U =
    \left( \begin{array}{ccc}  1 & 0 & 0 \\  0&\exp (-i \omega_1t) & 0 \\
             0 &0 &  \exp (-i(\omega_1-\omega_2)t) \end{array} \right )   \,,
\]
results in %\cmt{(this eq. needs to be checked)}
\[
\hat H^\prime = \begin{pmatrix}
    0 & \frac{1}{2}\vec d_{12}\left(\vec E_1 (e^{i\Delta_1 t}+e^{i(\omega_1+\omega_{21})t})+\vec E_2(e^{i(\omega_2-\omega_{21})}+e^{i(\omega_2+\omega_{21})t})\right) & 0 \\
    c.c. &  \Delta_1 & c.c.\\
    0 & \frac{1}{2}\vec d_{32}\left(\vec E_1 (e^{i(\omega_1-\omega_{32}) t}+e^{i(\omega_1+\omega_{32})t})+\vec E_2(e^{i\Delta_2t}+e^{i(\omega_2+\omega_{32})t})\right)& -\Delta_{2P}
\end{pmatrix}\,,
\]
where $\omega_{ij}=\mathcal E_i-\mathcal E_2$ are the transition frequencies, $\Delta_1=\omega_{1}-\omega_{21}$, $\Delta_2=\omega_2-\omega_{32}$ 
are the one-photon detunings, and $\Delta_{2P}=\omega_{32}-(\omega_1+\omega_2)$ is the two-photon detuning.
It is instructive to compare this Hamiltonian to the one obtained from transforming into the interaction picture, i.e., for $\hat U=\exp(i \hat H_0t)$ where the diagonal becomes zero and the detunings appear only in the off-diagonal matrix elements.

Now we assume that $\omega_{1}$, $\omega_2$ are near-resonant to the two transitions in the three-level system. This allows us to take the so-called two-photon rotating wave approximation where we neglect terms that oscillate with sum frequencies as well as those oscillating with the large difference frequencies $\omega_1-\omega_{32}$, $\omega_2-\omega_{21}$, while keeping those with the small difference frequencies, i.e., the one-photon detunings $\Delta_1$, $\Delta_2$. This reflects that while the electric field contains both spectral components, the probability for the first (second) component to excite transitions between $\ket{2}$ and $\ket{3}$ ($\ket{1}$ and $\ket{2}$) is very small. 
In other words, within the RWA each transition "sees" only the near-resonant component of the electric field. 
The Hamiltonian then becomes
\[
H^\prime_{RWA}=\begin{pmatrix}
    0 & \frac{1}{2}\Omega_1(t)e^{i\Delta_1t} & 0 \\
    \frac{1}{2}\Omega_1(t)e^{-i\Delta_1t} & \Delta_1 & \frac{1}{2}\Omega_2(t)e^{i\Delta_2t} \\
    0 & \frac{1}{2}\Omega_2(t)e^{-i\Delta_2t} & \Delta_{2P}
\end{pmatrix}\,,
\]
which is very often taken as the starting point~\cite{ShoreBook2011,VitanovRMP2017}, but one should keep in mind that it is the result of an approximation. 
%\cmt{to be added: the role of the one-photon and two-photon detunings, could consider  specific examples $E_2>E_3>E_1$ ($\lambda$-system), $E_3>E_2>E_1$ (ladder system), and $E_3>E_1>E_2$ (V-type system). }
     
Comparing the unitary transformations for two- and three-level systems, we can devise a general strategy for frame transformations and rotating-wave approximations for $N$-level systems: The unitary is diagonal and the complex phases determine by how much the corresponding lab-frame energy is shifted. In comparison to the interaction picture, where all lab-frame energies are shifted to zero and all detunings appear in the off-diagonal matrix elements of the Hamiltonian, one can also choose frame transformations where the detunings appear on the diagonal, and the off-diagonal matrix elements become real.

\end{document}